# iGEM: a model system for team science and innovation

Marc Santolini[1,2,3,7 †], Leo Blondel[1,2], Megan J Palmer[4,5 †], Robert N Ward[1,2,4,6], Rathin Jeyaram[1,2], Kathryn R Brink[5], Abhijeet Krishna[3], Albert-László Barabási[3,7,8]

1. Université Paris Cité, Inserm, System Engineering and Evolution Dynamics, F-75004 Paris, France
2. Learning Planet Institute, F-75004 Paris, France
3. Network Science Institute and Department of Physics, Northeastern University, Boston, MA 02115
4. Department of Bioengineering, Stanford University, Stanford, California, USA.
5. Center for International Security and Cooperation, Stanford University, Stanford, California, USA.
6. School of Public Policy, Georgia Institute of Technology, Atlanta, Georgia, USA
7. Channing Division of Network Medicine, Department of Medicine, Brigham and Women's Hospital, Harvard Medical School, Boston, MA, USA.
8. Department of Network and Data Science, Central European University, Budapest, Hungary.

[†] correspondences: marc.santolini@cri-paris.org, mjpalmer@stanford.edu

**Teams are a primary source of innovation in science and technology. Rather than examining the lone genius, scholarly and policy attention has shifted to understanding how team interactions produce new and useful ideas. Yet the organizational roots of innovation remain unclear, in part because of the limitations of current data. This paper introduces the international Genetically Engineered Machine (iGEM) competition, a model system for studying team science and innovation. By combining digital laboratory notebooks with performance data from 2,406 teams over multiple years of participation, we reveal shared dynamical and organizational patterns across teams and identify features associated with team performance and success. This dataset makes visible organizational behavior that is typically hidden, and thus understudied, creating new opportunities for the science of science and innovation.**

# Introduction

We increasingly rely on teams to produce new science and technology. Over 90 percent of science and engineering publications have multiple authors, more than half of US patents are co-invented, and in both cases the highest impact works are now team-authored[1]. These observations have prompted a flurry of research on the determinants of team innovation[2–4]. Yet, while large-scale metadata has made it easier to observe how composition (e.g. size, demographics, knowledge) affects success and other outcomes, our understanding of team structure is notably less advanced. Addressing this gap is key, since the way team members interact to perform their work ultimately mediates the effects of composition on their performance and success.

Team collaborations introduce a set of unique challenges, from communication to coordination, which, if left unaddressed, could jeopardize the completion and the success of the project. Prompted by the prevalence of the phenomenon, multiple studies have examined how team size[1,5], composition[6–11] and division of labor[12–16] shape innovation, usually relying on conceptual models[17] or proxies from the metadata associated with journal articles, patents and grants[18]. They often benefit from large samples but are limited in the granularity and scope with which organizational activity can be observed. While the CRediT contributor roles taxonomy is increasing in adoption, most of the available data is difficult to parse and comes with accuracy concerns[19]. Meanwhile, qualitative studies [20–22] provide rich detail at the expense of scale and scope. Finally, while surveys can be run at scale[23–26], low response rates are common, repeated measures are difficult to find, and data often remains proprietary to its creators, limiting follow-on work. Therefore, despite a flood of research on team science and innovation, issues with data access and utility have limited the extent of our understanding of the micro-level collaborative processes underlying performance.

Responding to this need, we introduce a large-scale panel dataset to study the organization of team science and innovation: the international Genetically Engineered Machine (iGEM) competition in synthetic biology (Fig 1)[27–30]. The utility of iGEM as a model system rests on three features. First, iGEM teams resemble those in professional settings in organizationally relevant ways. They work in a complex environment with high task interdependence and uncertainty[31,32]. They are largely self-assembled within institutional and geographic constraints[9,33], of similar size[1] and structure. They work towards providing proof of concept for scientific or technological innovations as evaluated by panels of experts in their fields. Second, competition rules require that teams report fine-grained data on their projects that cannot be found in the publicly available article data widely used today. This includes static and dynamic behavior, internal (within team) and external (across team) interactions, as well as text describing the team and how its project came to be. Finally, while teams often participate multiple years in a row, the competition rules have remained largely stable, supporting the use of panel data estimators for causal inference.

In this study, we describe the unique features of iGEM as a model system for team science and innovation, and showcase how the iGEM data can be used to observe team behavior and to identify drivers of team performance, success and learning over time.

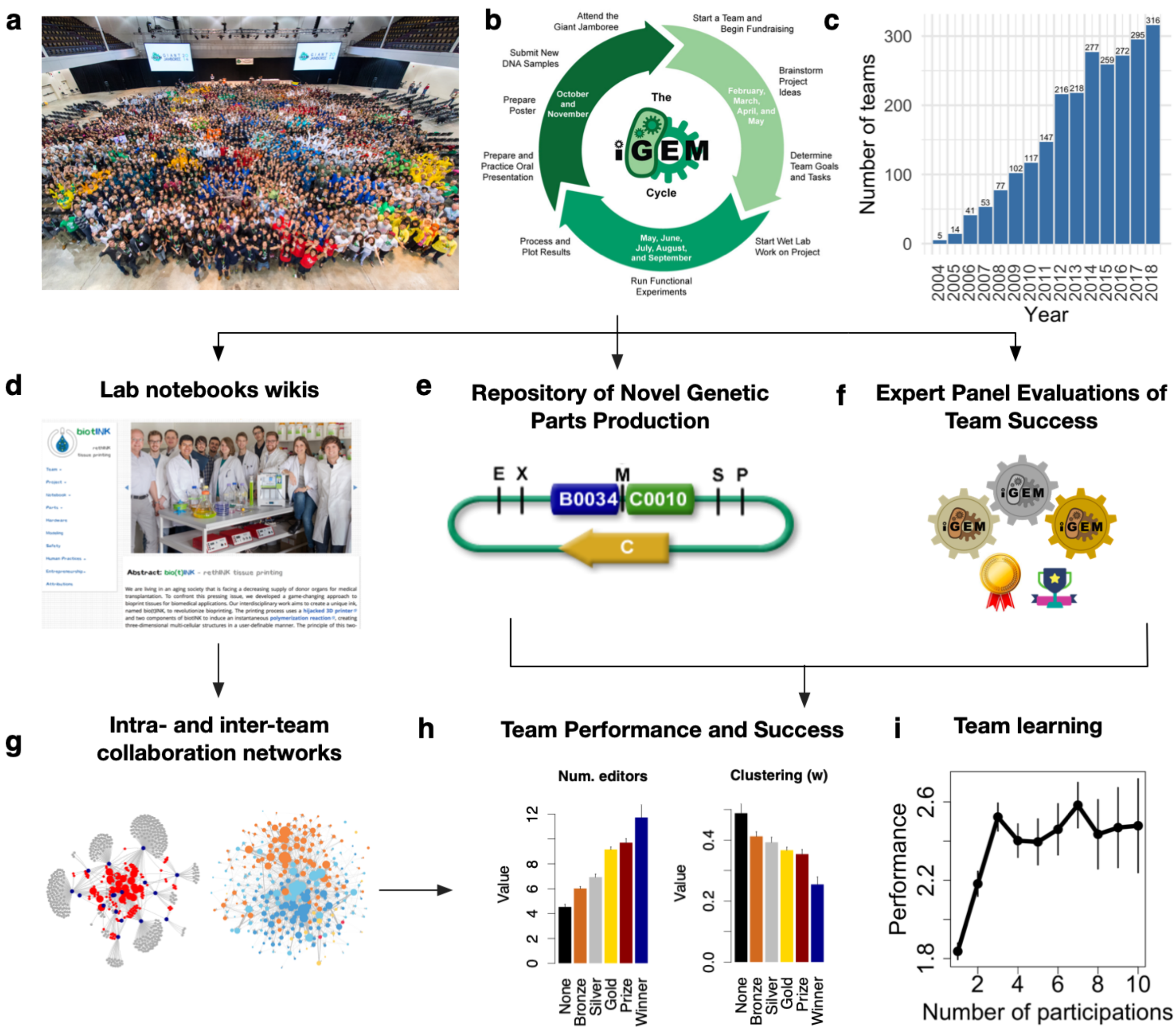

**Figure 1 - iGEM: an annual international team-based science and engineering competition**

**a.** Picture taken at the 2014 iGEM Giant Jamboree in Boston, showing attending teams. **b.** The official iGEM competition cycle, showcasing the workflow of team work during the competition time, from May to October. **c.** Number of teams per year across competitions, showing a steady increase since the start of the competition. **d.** An example wiki page of an iGEM team, displaying team information and project results. **e.** Illustration of a BioBrick, a genetic part produced by teams. **f.** Team results consist of medals and awards. **g.** Team networks can be derived at the intra-team level (left) and inter-team level (right). The intra-team bipartite network is derived from wiki contributions, and the inter-team network from mentions in the wiki. See Fig 3 for more details. **h-i.** Using performance data across multiple team participations, we uncover features associated with performance (h) and learning across multiple participations in the competition (i). Panels b and e were obtained from the iGEM website[34].

# Results

## iGEM is a unique model system for studying team innovation

Each year the international Genetically Engineered Machine (iGEM) Competition challenges teams to create new innovations in synthetic biology (Fig 1). Since its foundation in 2004, the competition has grown from 5 to hundreds of teams across 40 countries. As of 2018, 2,406 participating teams and over 30,000 team members had taken part (Fig 1c and Fig S1) [28]. The competition has been a global catalyst for the synthetic biology field: many alumni have started their own academic labs and as of 2021 have founded 175 start-ups in over 40 countries[35]. In line with the growth of iGEM, synthetic biology is now recognized in the US and other countries as a national innovation priority [36,37].

The competition runs each year (Fig 1b). Teams assemble and plan their work before the official registration date in May, when they receive a standard set of biological parts—BioBricks—to build their projects with. Their challenge is to design and build synthetic biology solutions to real world problems. While teams work with the same parts and are evaluated on the same criteria, they have broad discretion in project choice and in how they conduct their work. Teams work through late October, culminating in a “Jamboree” where they congregate to present the fruits of their labor. Team performance is evaluated by panels of judges based on written documentation, a poster and an oral presentation according to a common and publicly available rubric (see details in SI). While competitive in nature, teams are incentivized to collaborate as part of the medal requirements. Aggregated evaluations determine non-competitive awards for absolute performance (gold, silver and bronze medals) as well as competitive awards for best in class (for an application area or particular element of the competition) and best in competition within each division (high school, undergraduate, overgraduate).

Most important for us, the competition rules enforce high quality documentation of team work. Each team produces a wiki during the competition cycle, a website that serves as historical documentation and the primary basis judges use to evaluate their work (Fig 1d). Wikis describe team composition, project design, tasks—from laboratory experiments to mathematical modeling and stakeholder engagement—, collaborations and more. Because these documents are collaboratively produced, patterns of contribution (i.e. who documented what) provide further data on team work across tasks and time. For each year in our data, iGEM hosts a Wikimedia wiki instance, allowing anyone to access the full history of revisions made by individual participants, which we refer to as edits. These individually identified, time-stamped edits provide a digital trace of team behavior as it unfolds throughout the competition cycle. In addition to their project wiki, a public team information page reports metadata about the team, including team roles (instructors, student leaders and members), and information about the new BioBricks they produced.

Linking the various datasets generated by the iGEM rules provides a richer, more dynamic view of team work than exists in publicly available datasets today. While these teams exhibit many of the same organizational features as professional science and engineering teams, iGEM teams work in public, therefore making it possible to examine behavior that would otherwise be hidden to protect competitive interests. In the following sections we highlight a few of the ways this data can be used to understand team organization, performance and learning.

## Stability of the competition ecosystem

Part of what makes iGEM useful as a model system is that teams participate multiple times over successive years within a largely stable set of rules. This section describes team and competition-wide behavior that has remained stable over the timespan of our data, while the supplemental information and Figures S2-3 detail the few competition rule changes across the same period.

## Team composition, productivity and performance

The data was obtained through automated content extraction from the official iGEM competition guidance websites and team wikis (see Methods). The resulting dataset covers team composition, wiki edit history, new BioBrick contributions and project success in terms of medals, prizes and awards (see Fig 2 and Methods).

A total of 2,406 teams participated in the competition between 2004 and 2018 (Fig 1c). While initially more centered on the North American region, the competition has seen a steady growth in representation from the Asian continent (Fig 2a and S1). Most teams are based at educational institutions or at community labs. Beginning in 2013, each team competes within a Division that corresponds to the academic level of their members: high school, undergraduate or overgraduate (Fig 2b). Team participants are grouped into hierarchical roles that distinguish between the students who design, implement and present the project, and mentors (PIs, advisors and instructors) who provide guidance and resources such as laboratory equipment (Fig 2c). Divisions and roles have remained unchanged since 2013 (Fig 2b,c).

Starting in 2008, each team was provided a template wiki with pages and sections to document different parts of their work. We extracted 2,577,845 wiki edits from 13,654 editors across 2,278 teams which participated in the 10 competitions between 2008 and 2018, resulting for each edit in a unique editor ID, a timestamp, a page and section edited, and a signed edit size (see Methods). Team sizes, not regulated by the competition, widely vary (Fig 2d and S5).

The average team makes 1,130 edits during the competition cycle, a number that is stable across 10 years of the competition (Figure 2e). Edit size (the amount of content) is distributed log-normally, with more than two times more positive (additions) than negative (removal) edits (Fig S7). In most teams, edits are made by a subset of team members, the wiki editors, that we call a team's active core (see Supplemental Information). The average team has about 7.4 wiki editors, but some have as many as 37. The size of the active core scales with the overall team size (Fig S6) and is more strongly associated with measures of performance and success than the overall team size (Fig S10).

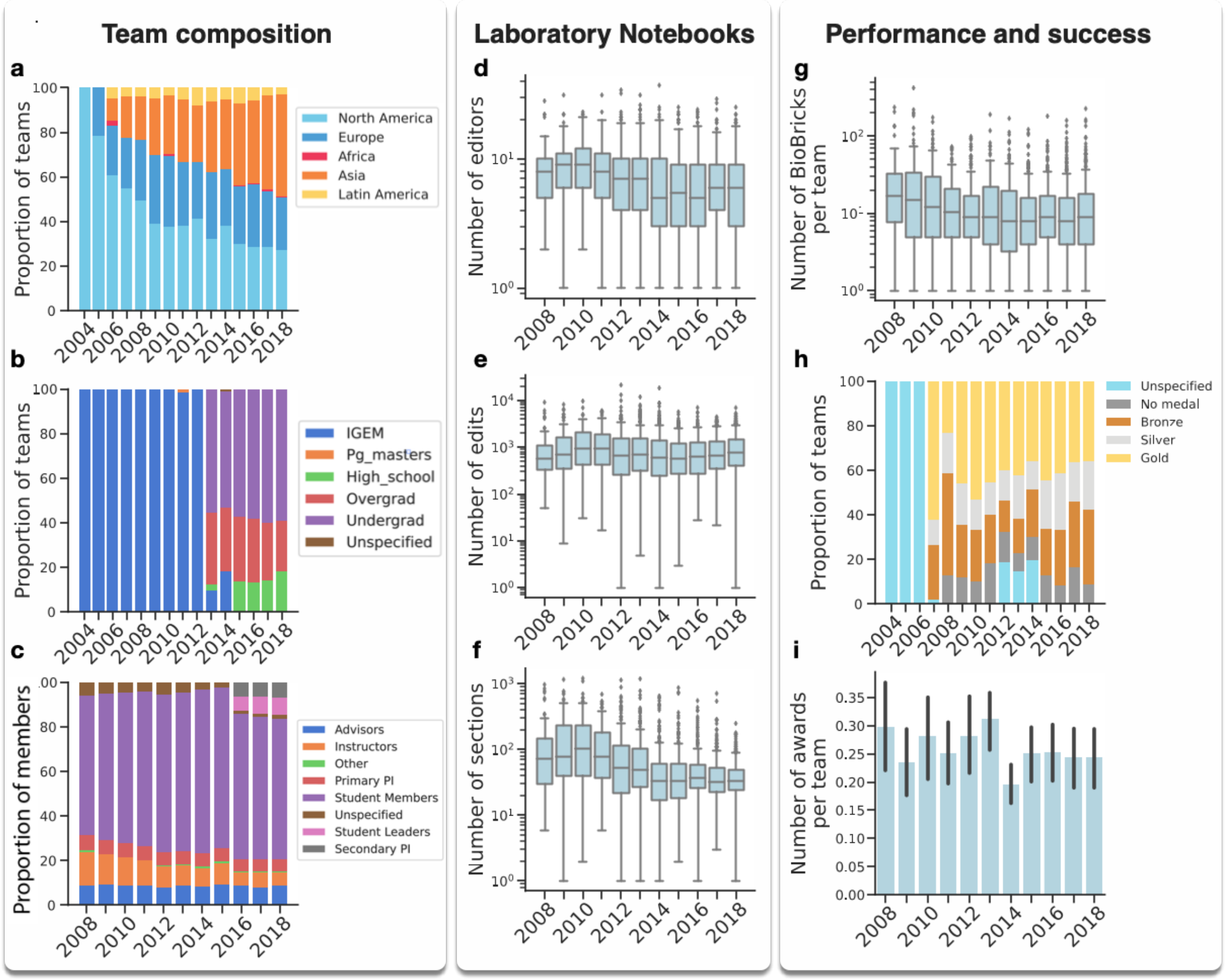

**Figure 2 - Open data on team composition, production and performance**

**a**. Geographical distribution of participating teams per year. Geographical provenances are binned into Regions by the iGEM headquarters. **b**. Distribution of team educational levels per year, called Divisions. Prior to 2013, only undergraduate teams participated in iGEM. Starting in 2013, teams competed in divisions corresponding to the maximum educational level of teams. **c.** Distribution of team members roles per year, an information available starting in 2008. In 2016, iGEM created two roles: Student Leaders and Secondary PI **d-e** Distribution of participants which committed at least one edit on the wiki (“editors”, **d**), of the number of wiki edits (**e**) and of wiki sections (**f**) created on the wiki pages per team across years. Boxplot whiskers extend up to 1.5 times past the low and high quartiles, after which data points are shown as outliers. **g.** Number of BioBricks produced by a team each year. **h.** Distribution of types of medals delivered to teams across years. **i.** Average number of prizes received by a team across the years. Error bars denote standard error.

Finally we extracted the number of new BioBricks produced by each team (Fig 2g) and the competition results from the official competition website. Teams earn medals (none, bronze, silver, or gold) by fulfilling requirements, and special prizes or awards by performing better than others on particular dimensions (Fig 2h,i). The proportion of teams earning medals, prizes and awards is roughly constant since 2008 (Figs 2h,i, S8), in part due to the growth in the number of available prizes per team (see Supplementary Information and Fig S3).

## Dynamics and structure of the team ecosystem

**Edit dynamics and deadline effect.** Timestamped wiki edits show how teams document their work throughout the competition. Projects officially start with team registration in late March (or early April) and run until a deadline in October when wiki content is frozen and begins to be used for evaluation (Fig 1b). Teams start editing as early as 200 days (~7 months) before the wiki freeze, rise to a plateau of high activity over the summer, before a final rush until the wiki freeze deadline (Fig 3a), a behavior consistent across competition years (colored lines and Fig S11). The probability of making an edit before wiki freeze time $T^{*}$ diverges as $P(t) \sim (T^{*} - t)^{-\alpha}$ with $\alpha = 1.06 \pm 0.07$, stable across competitions (Fig 3b,c) and consistent with a deadline effect[38].

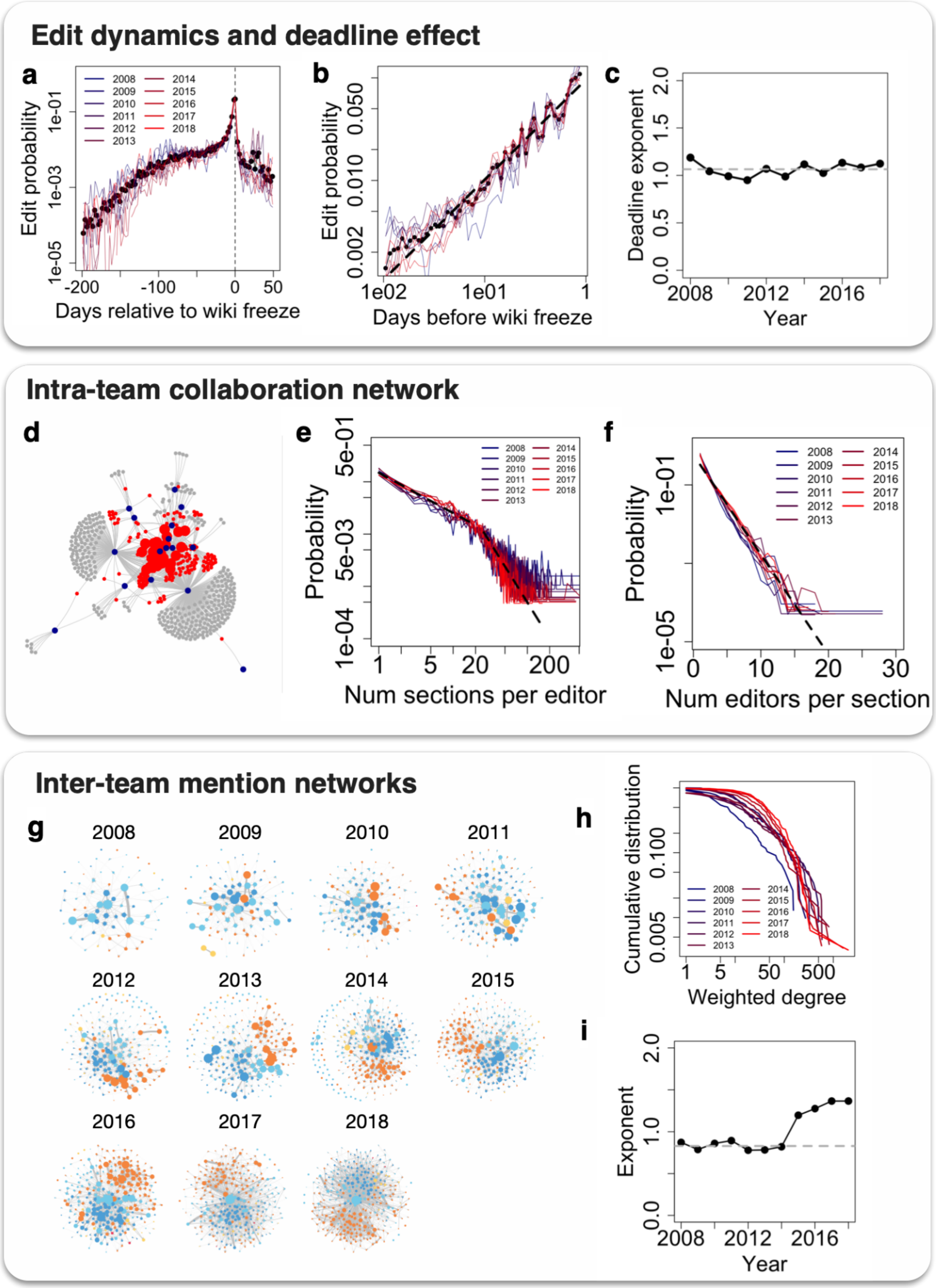

**Figure 3. Dynamics and structure of the team-of-teams ecosystem**

**a.** Temporal distribution of wiki edits as a function of time to wiki freeze, per competition cycle. **b.** Same as a., in log-log space. Dashed line shows linear regression for the last 30 days before wiki freeze. **c.** Exponent from the linear regression model of b. as a function of competition cycles. **d.** Example team bipartite network, derived from wiki contributions. Blue nodes correspond to team members, gray nodes are wiki sections (see Methods) edited by a single team member, and red nodes correspond to collaborative sections edited by multiple team members. Wiki sections edited by more than 3 team members are shown with a larger node size. **e.** Distribution of the number of sections per editor, per competition cycle. Dashed lines correspond to regression models for the 0-20 and the 25-150 regions. **f.** Distribution of editors per section, per competition cycle. Dashed line shows a regression model for the 0-20 region. **g.** Inter-team mention networks across competitions. A directed link is drawn between (source) teams mentioning other (target) teams in their wiki, and are weighted by the number of mentions (see Methods). Colors correspond to Regions: Africa (red), Latin America (yellow), North America (light blue), Asia (orange), Europe (blue). **h.** Receiver-Operator Curve (ROC) analysis of the number of mentions in a wiki as a predictor for a collaboration between two teams, using a manual annotation for the 2015 network. **i.** Complementary cumulative distribution of normalized collaboration strength across years, in log-log scale. The normalized strength is defined by the ratio of the weighted degree of a node (sum of all incoming and outgoing edge weights) to the average strength for each year. Curves collapse to a single collaboration distribution. **j.** Stability of the exponent of a power-law fit to the strength distribution across years

**Team organization.** In addition to the time spent on the project, team members need to divide labor, coordinate and integrate the results of multiple tasks in the course of their project. Wikis are subdivided into sections that describe different aspects of the project. We use the individually identified edits made in each section to infer which members worked on which aspects. We represent the division of labor in each team as a bipartite network connecting team members with the wiki sections they edited (Fig 3d). The number of sections per editor in this network—a proxy for the number of tasks editors performed[39]—has a heavy-tailed distribution that decays slowly up to 20 sections, and falls abruptly after (Fig 3e and S12a). On the other hand, the number of editors per section, an indicator of collaboration, follows a bounded, exponential distribution (Fig 3f and S12b). The number of editors in a particular section rarely exceeds 8, suggesting a characteristic subteam size that is conserved across teams. While some sections are written collaboratively (large red nodes, Fig 3d), most sections are written by a single team member, a behavior conserved across competition cycles (Fig 3f).

**Inter-team collaborations.** Beyond internal collaboration, iGEM teams are incentivized by medal requirements to collaborate with other teams. Much like professional settings, teams exchange materials and advice[24,40], solve problems[41] and socialize together[42]. The incentive for collaboration increased when "helping another team" changed from a gold to silver medal requirement in 2016 (Fig S2). Along with a 2015 increase in the number of requirements teams have to satisfy to earn a gold medal from one to two, collaboration within the competition has increased as a whole (Fig 3g).

Teams describe their interactions in natural language on their wiki, making it possible to filter the collaboration network for specific questions (e.g. about in-person interactions, or advice giving) in a way that cannot be done with most existing datasets[43,44]. We construct a directed, weighted network for each competition year, by parsing each interaction mentioned on each team wiki; edges are weighted by the number of times a peer team is mentioned by the focal team, a measure of tie strength (Fig 3g). To validate this measure we leveraged an independent database of "significant" team collaborations from the 2015 competition that was manually constructed by the 2016 iGEM Waterloo team for their project[1]. We performed a Receiver-Operator Curve (ROC) analysis to understand whether the number of team mentions in a wiki predicts significant collaboration between two teams as assessed by Waterloo (Fig S13). The retrieval rate is high, AUC = 0.87 (Wilcoxon $p < 2.2e\text{-}16$), supporting the use of wiki mentions to measure significant collaboration between teams.

[1] http://2016.igem.org/Team:Waterloo/Integrated_Practices/Networks

The complete inter-team networks we obtain (containing edges of any time) are assortative on the basis of geographic location (Fig 3g). Their degree distributions are skewed with a few high degree teams and a heavy-tail containing many less connected ones (Fig 3h). We observe an increase in the decay exponent of these distributions in 2015 and 2016 after iGEM increased the incentive for collaboration. This rule change increased the minimum weighted degree we observe while leaving the maximum weighted degree unaffected (Fig 3h).

Altogether, these results show that team behavior is largely consistent over time, but responsive to a few major rule changes, such as the collaboration requirement introduced in 2015 (Fig 3i). Teams contribute to their wiki project throughout the competition period, with a rush to the deadline. The span of contribution varies widely, with a minority of editors working on very large numbers of wiki sections. Finally, we find that the number of individuals collaborating on a wiki section is approximately bounded at 8 members, possibly to reduce coordination costs[31,45]. On the other hand, external division of labor—the number of external collaborations a team engages obeys no such bound[25].

## Shared patterns of intra-team organization

Next we explore how basic organizational features—team size, division of labor, and the dynamics and distribution of effort[32]— vary within our data.

**Team dynamics.** In order to achieve a collective goal, team members need to synchronize their activity, a behavior foundational to team cooperation[32,46] and collective intelligence[8,47]. We showed in Fig 3b that overall wiki edits follow a deadline effect $P(t) \sim (T^{*} - t)^{-\alpha}$ at the competition scale. This observation also holds within single teams (Fig 4a), with $\alpha = 1.02 \pm 0.77$ stable across team sizes (Fig 4b, N=2,011 teams with more than 100 edits), indicating that members of different teams have common posting routines as the deadline approaches.

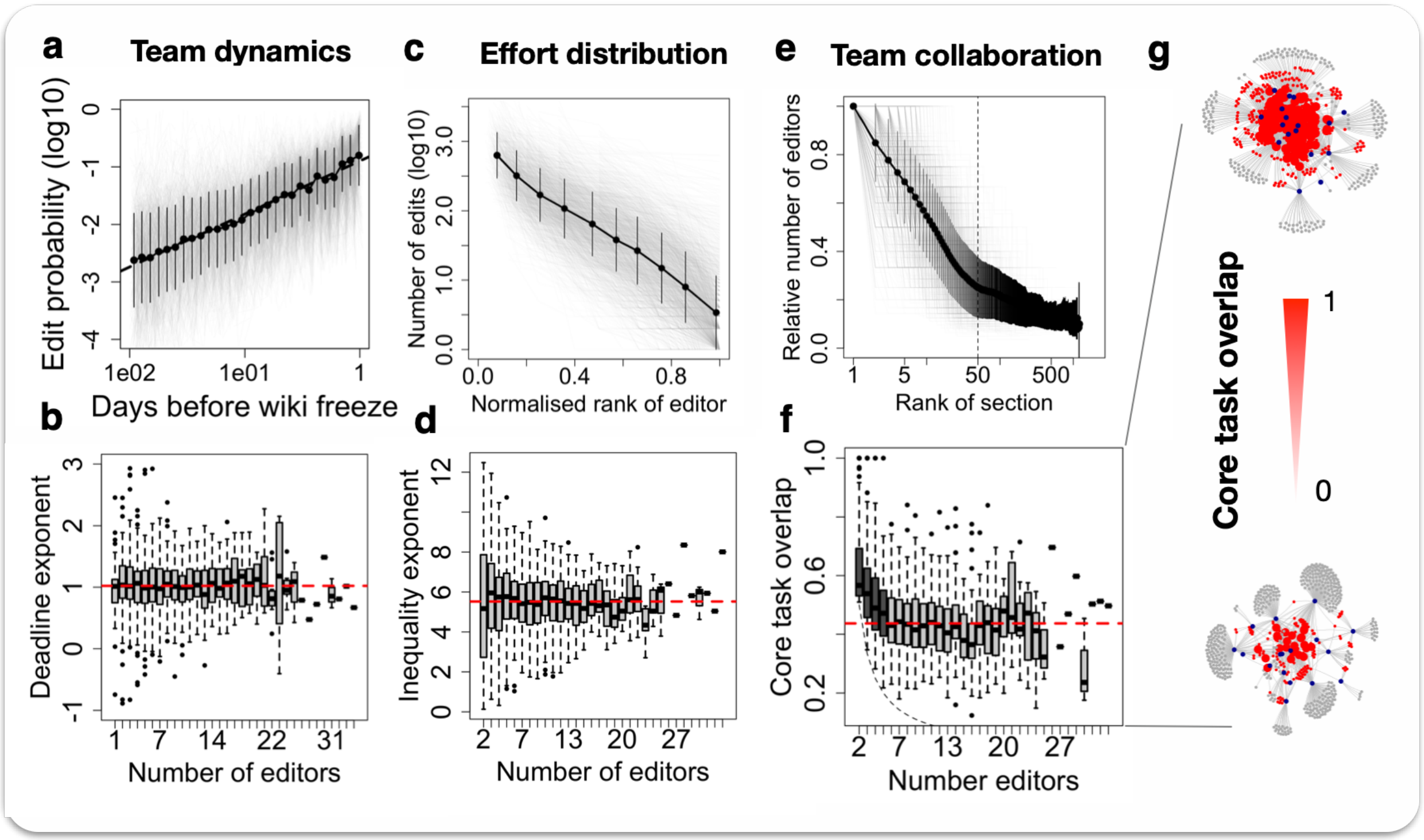

## Figure 4. Shared patterns of teamwork in the competition

**a.** Temporal distribution of wiki edits as a function of time to wiki freeze, in log-log space. Gray lines show individual teams, and the black line shows average and standard error across teams. Dashed line shows linear regression. **b**. Distribution of the exponent from the linear regression model of a. for individual teams, as a function of team size. We find consistent values across team sizes (p=0.77, Spearman correlation). **c.** Number of edits (Log10) as a function of the proportion of team members, sorted by decreasing contribution. Gray lines correspond to individual teams, and the black line shows the average and standard deviation across teams. Linear trends indicate exponential distributions. **d.** Best-fit exponents to the exponential distributions of a. as a function of team size (p=0.15, Spearman correlation). **e.** Relative number of editors per section, ranked by decreasing number of intra-team collaborators. Gray lines correspond to individual teams, and the black line shows the average and standard deviation across teams. Teams fall under a similar curve, with a decay until 50 sections and a flattening afterwards. We refer to the top 50 sections as the "collaborative core". **f.** Core task overlap, corresponding to the area under the curve before 50 sections in e, as a function of team size. The black dashed line corresponds to the minimum possible values of $1/N_{editors}$, creating a distorsion for very small teams (size < 5). The core task overlap for teams of size larger than 5 are stable (p=0.12, Spearman correlation). The red dashed line indicates average core task overlap for teams larger than 5. **g.** Example bipartite networks with high (top) and low (bottom) core task overlaps.

**Effort distribution.** All organizations face a problem of motivating members to expend effort[32,48]. Studies of open-source communities suggest that in high-performing teams, a few active contributors[12] do most of the work. Here we use the number of wiki edits per editor to proxy the effort given by individual members to their team. To characterize the distribution of effort within a team, we rank editors by decreasing number of edits, and computing a "normalized rank" $\widehat{r_i} = (r_i - 1)/(N - 1)$, where $r_i$ is the rank of editor $i$ in the team ordered by decreasing number of edits, and $N$ is the number of editors in the team. The most active editors on a team have $\widehat{r_i} \simeq 0$ while the least active have $\widehat{r_i} \simeq 1$. Individual effort (number of edits) as a function of normalized rank follows $N_{edits} \sim e^{-\alpha r_i}$ (Fig 4c), with exponent $\alpha = 5.5 \pm 1.8$ conserved across team sizes and competition cycles (Fig 4d and S14). In other words, most of the editing is performed by just 20% ( $1/\alpha$) of the team.

**Division of labor.** Team projects require their members to handle tasks ranging from fundraising and wet lab biology, to software engineering and mathematical modeling. The diversity of tasks and volume of work to be done suggest that teams can find performance gains through division of labor[45,49]. We examined how teams allocate members to tasks using the same editor-section bipartite network used above (Fig 3d). When the division of labor is low (high), we expect to see many sections that are edited by most (few) members. We rank the sections of a team's wiki by the number of editors that worked on them, normalized to the section with the most editors. The top ~50 sections are edited by multiple team members, after which many less collaborative sections are edited by a single editor (Fig 4e and S15). To quantify the level of collaboration in top sections, we define the *core task overlap* for a team as the area under the curve in Fig 4e for the top 50 sections. The higher the core task overlap, the larger the proportion of team members involved in working on the top sections rather than on more peripheral sections, i.e. the lower the division of labor. Fig 4g shows teams with a similar number of team members and wiki sections, that exhibit high versus low core task overlap. The top sections are visualized as red nodes. Their size indicates the number of members that edited them. Thus red nodes will be larger when core task overlap is high. When only one section is collaborative and all others are peripheral, we obtain a lower bound for the core task overlap of 1/N, where N is the number of editors; but when N is sufficiently large (5+ editors), this lower bound is vanishingly small and the core task overlap converges to a constant median value (Fig 4f).

Taken together, these results show that the deadline rush, contribution inequality and bounded subteam sizes exhibited at the aggregated level in Fig 3 are stable features across teams despite their vastly different sizes, showcasing the homogeneity of the patterns of teamwork within the iGEM ecosystem.

## Team performance and success

Finally, we show how the richness of iGEM data, and variation in team characteristics can be used to explain performance and success in the competition. We focus on three outcomes: the medal a team received (none, bronze, silver, or gold), whether they were awarded a special prize, and whether they won the competition, organized along a performance scale of 1 to 6 (Fig 5a-e). We then highlight associations between the outcomes of a team and their composition (Fig 5a), wiki editing behavior (Fig 5b), local and global position in the inter-team network (Fig 5c) and experience from previous competitions (Fig 5d).

The most fundamental aspect of team composition is the number of people working on the project. Because teams often have registered members who make minimal contributions, the number of students in the team is a weaker predictor of success (Pearson correlation $r=0.08$, $p=3e-4$) than the number of wiki editors ($r=0.35$, $p=5.6e-60$), confirming that the latter is a more relevant measure of an active team core (Fig 5 a-b). On the other hand, we find that the number of advisors ($r=0.15$, $p=1.7e{-12}$) and instructors ($r=0.22$, $p=1.1e{-24}$) are both positively associated with increasing success. This is consistent with several possible mechanisms that beg further investigation, including the speed and individual specificity with which team members receive feedback, and the breadth of tasks their mentors can provide guidance on.

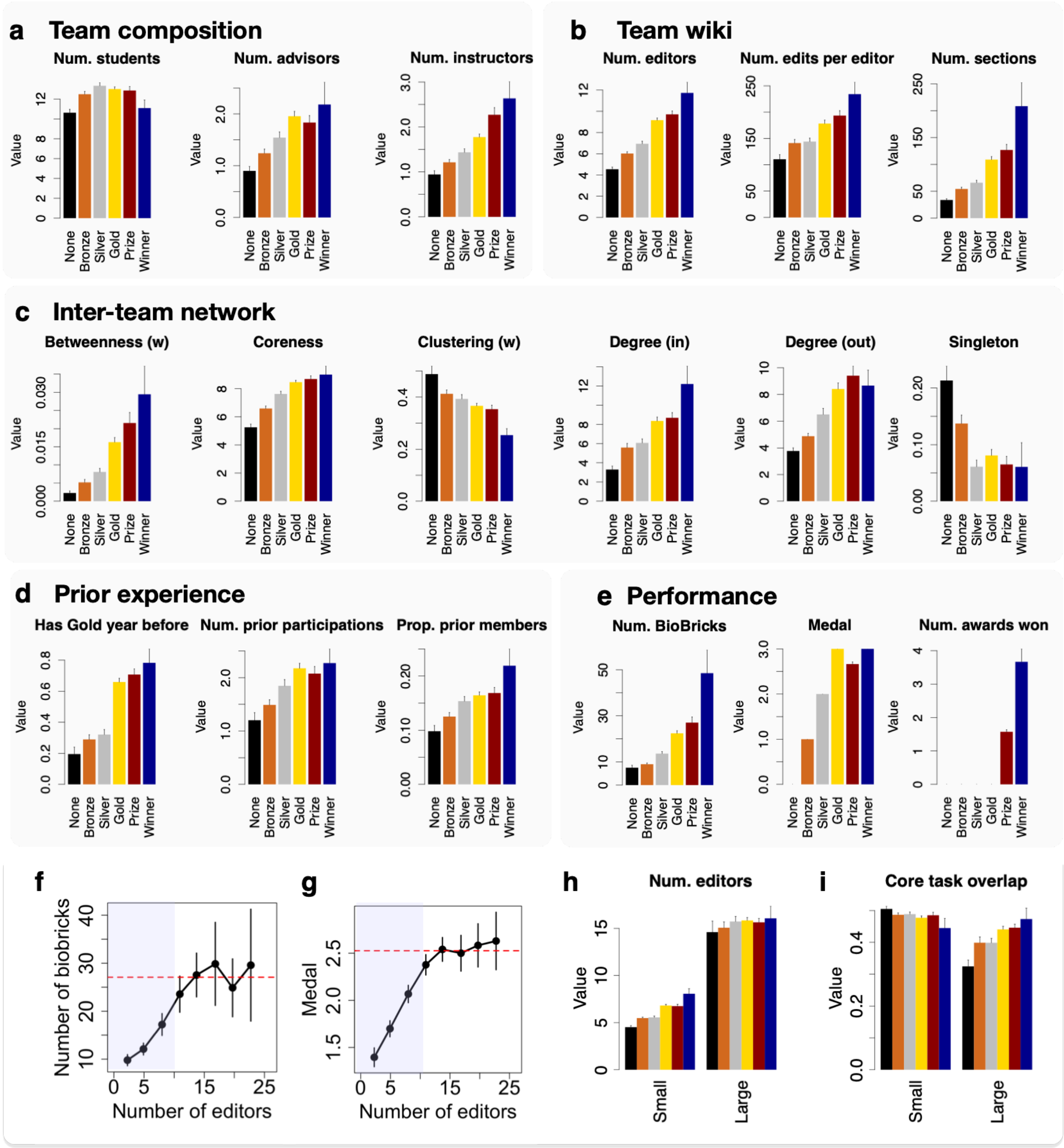

**Figure 5. Team performance and success**

**a-e**. Barplots showing average values and standard errors for various team features related to composition, documentation, collaborations, experience and performance across competition outcomes corresponding to increasing medal categories (No medal, bronze, silver or gold medal), followed by the obtention of a special prize, and winning the competition. Note that special prizes can be obtained even if no gold medal was obtained. **f.** Number of BioBricks, a quantitative marker of team productivity that is not explicitly incentivized in the judging criteria, that only require a single novel BioBrick. Dashed line shows the saturation level for teams larger than 10. **g.** Same as **f**, for Medal, assessed using judging criteria evaluating project quality, and quantified here using a variable taking values 0 (no medal), 1 (bronze), 2 (silver), 3 (gold). **f.** Same as in a-e, grouping teams into small (<12 editors) and large (12+) teams.

Wiki edits proxy the volume and structure of content a team creates, and are thus important for performance and success (Fig 5b). Beyond the number of editors involved in the wiki, successful teams achieve higher individual engagement in wiki edits (r=0.22, p=3e−23) and produce a larger number of wiki sections (r=0.27, p=3.4e−36), an indicator of a more detailed project.

By collaborating with peers, teams access knowledge, materials and support that may help their performance[40]. In particular, the structural position of a team in the collaboration network is likely to affect the resources a team can access, the willingness of their collaborators to provide help, and the ease with which they can apply external resources in their own project[50,51]. In iGEM, we find that performance is positively associated with both local and global centrality measures (Fig 5c). Indeed, we find that the number of direct neighbors is important for success, as quantified by the number of mentions from (in-degree, r=0.19, p=1.5e−18) and towards other teams (out-degree, r=0.2, p=1.4e−19), as well as the absence of ties in the collaboration network (r=−0.15,

p=1.9e−11). This may reflect the fact that showcasing at least one significant collaboration is a medal requirement with an increased incentive (silver medal) in later years. Beyond the local neighborhood, we find a positive association of performance with the depth of embeddedness of a team in the network, measured by global centrality measures such as betweenness (r=0.22, p=8.1e−25) and coreness (r=0.29, p=6.7e−42) centralities. Finally, we find a negative association with clustering (r=−0.13, p=2.6e−09), indicating that successful teams tend to be in structurally open network positions associated with access to diverse resources and perspectives[50].

Finally, we find that prior experience (number of prior participations, r=0.13, p=6.2e−10, proportion of team members who participated to previous editions of the competition, r=0.14, p=6.9e−11) and recent success (having obtained a gold medal the previous year, r=0.37, p=2.8e−36) strongly predicts present success (Fig 5d), consistent with the success-breeds-success dynamics common in other settings, a phenomenon which we explore further in the next section.

The quality of a team's project is a result of their performance on multiple tasks. Larger team sizes are advantageous, as reflected in the positive association between number of wiki editors and performance (Fig 5b). However, as teams grow and specialize, it may become harder to coordinate work and integrate outputs[45,52]. While the number of wiki edits per capita, a measure of individual member engagement, is stable across team sizes (Fig S16), the number of BioBricks produced and the medal (i.e. level of success) achieved both show nonlinear behavior: linearly increasing for small teams followed by a saturation for teams larger than N≃12 editors (Fig 5f,g, stable across competition cycles in Fig S17). As such, while adding more members may increase the performance of small teams, it ceases to be an effective strategy for large ones (Fig 5h, r=0.3, p=8e-34 for small teams N < 12, r=0.05, p=0.32 for large teams with N ≥ 12 members). Inversely, we find that the ability of team members to collaborate on core tasks (higher core task overlap) shows a strong positive relationship with performance for large teams of N ≥ 12 members (r=0.2, p=4.2e-5), while showing only a weak association for small teams (r=0.07, p=0.03 after restricting to teams with N > 5 to avoid the saturation effect from Fig 4f).

Altogether, this analysis suggests multiple strategies for recruitment and organization to improve team performance and success. Teams need to motivate members to contribute, structure their contributions and integrate their outputs to succeed[32,53]. These problems become harder with increasing size as coordination costs mount [31,45,54,55]. The median team size is 11 while the median number of editors is 6. Our results suggest that either recruiting more members or motivating greater individual engagement would substantially increase the performance of most teams. In addition, we find that teams with experiences of recent success, strong supervision, access to diverse collaborators, coordinating the members of the team over core project activities, and demonstrating high engagement generally perform well. Overall, the ability to measure each of these dimensions across many teams in the same environment provides new opportunities for understanding the mechanisms that cause the associations we report here.

## Dynamics of team success, lock-in effect and recovery strategies

Learning and improvement over multiple projects and years is a central concern in organizational theory and practice. Yet, most studies of team science and engineering take a cross-sectional perspective, focusing on data gathered at a specific point in time rather than tracking changes over time. One of the most useful aspects of the iGEM data is that it follows the same teams over multiple years with all of the granularity described above. As shown in Fig 5, the ability to capitalize on past success underlies team performance and success, a finding in line with the idea that early success begets success at the individual level in terms of scientific funding and recognition[56–58]. Here we study teams over multiple years of participation to understand the dynamics of team learning and success.

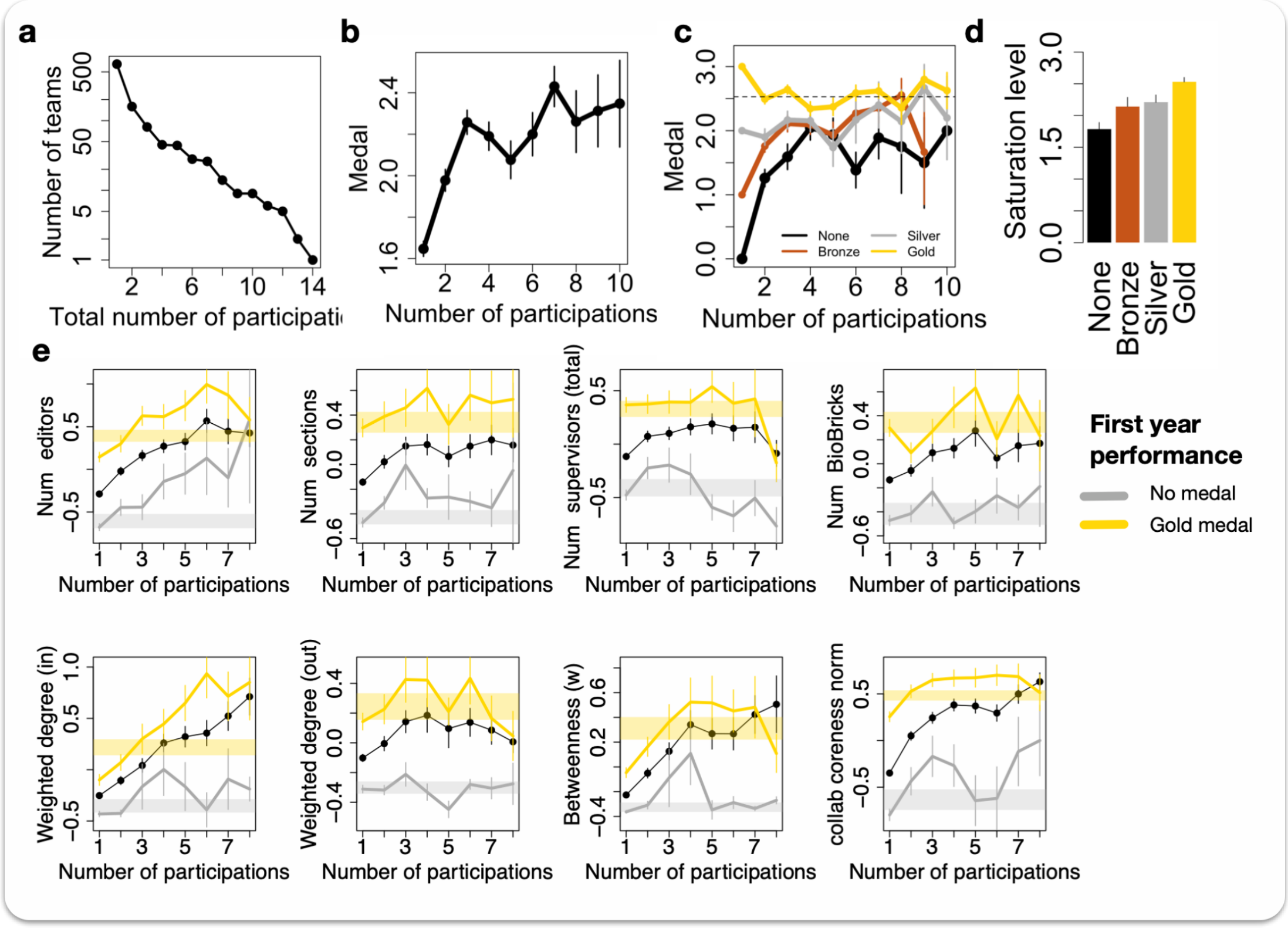

## Figure 6. Long-term dynamics and lock-in effect

**a.** Number of teams as a function of their total number of recorded participations in the competition. **b.** Average medal obtained as a function of participation count. As in Fig 5e, medals are quantified using a variable taking values 0 (no medal), 1 (bronze), 2 (silver), 3 (gold). Error bars denote standard errors. **c.** We split teams into 4 groups depending on their outcome during their first participation: teams which did not get a medal (black), teams which got a bronze (brown), silver (gray), or gold (yellow) medal. We show the medal obtained by these groups as a function of the number of participations. Dashed line corresponds to the average medal obtained by initial gold medalists after 3rd participation (saturation level, see d). **d.** Teams with an initial setback (no medal) have a significantly lower saturation level after the 3rd participation (average and standard error across relevant years), indicating a lock-in effect (p=6.5e-5, t-test). **e.** Evolution of Z normalized features across participations (see Methods). We showcase two subgroups of teams with an initial setback (no medal on first participation, gray) and an initial success (gold medal at first participation, yellow). Shaded areas correspond to mean -+ 2 standard errors of the feature value across gold medalists (yellow), and across teams which did not get a medal (gray), irrespective of the competition year.

**Longitudinal analysis.** Of the 1,075 unique teams (i.e with a stable team name across competitions, see Methods) that participated in the competition, 429 (40%) have participated more than once (Fig 6a), with a median of 1 prior participation per team across competition cycles after 2010 (Fig S18). On average, the probability of team success increases over their first 3 participations, saturating thereafter (Fig 6b). This trend can also be observed in the dynamics of team composition (number of editors, number of supervisors), productivity (number of wiki sections, number of BioBricks), and integration in the inter-team network (weighted degree, betweenness centrality, and coreness), as shown in Fig 6e. This suggests that some teams learn to organize more effectively over successive participations, allowing them to reach a plateau of consistent success.

**Lock-in effect**. While average success increases over repeated participations, not every team improves: we observe a strong impact of the performance at first participation. Initially low performing teams (no medal or bronze) strongly improve during the first 3 participations before performance saturation, while initially high performing teams (silver and gold) keep a similar performance over this period (Fig 6c). But these groups do not regress to the same mean. Indeed, we find that the saturation level of success in the long term depends on whether the team experienced failure upon first participation: teams with an early setback (no medal the first year) show a 30% lower success in the long term compared to teams with an initial gold medal (Fig 6d).

This suggests a lock-in effect where success in the first year partly determines future team success and the way their composition, behavior, and productivity changes in time (Fig 6e). Teams with different initial conditions (no medal vs gold medal in the first year) show a sustained or increasing gap in their characteristics over multiple participations. In particular, the weighted degree of incoming collaborations, a proxy for the extent of team involvement in the inter-team network, exhibits an increased gap over time, which might reflect a preferential attachment mechanism where teams with a significant prior success are able to attract more collaborators.

**Recovery from an early setback**. Still, some teams that experience an early setback do escape the lock-in effect. Out of the 160 teams with no medal or a bronze medal in their first participation, 73 (45%) eventually receive a gold medal in the competition (Fig 7a, red curve). These teams are able to match their success to the average success level at the second year of participation. Starting with the same initial success, we find 87 teams that never achieve better than a silver medal throughout their re-participations (Fig 7a, gray curve).

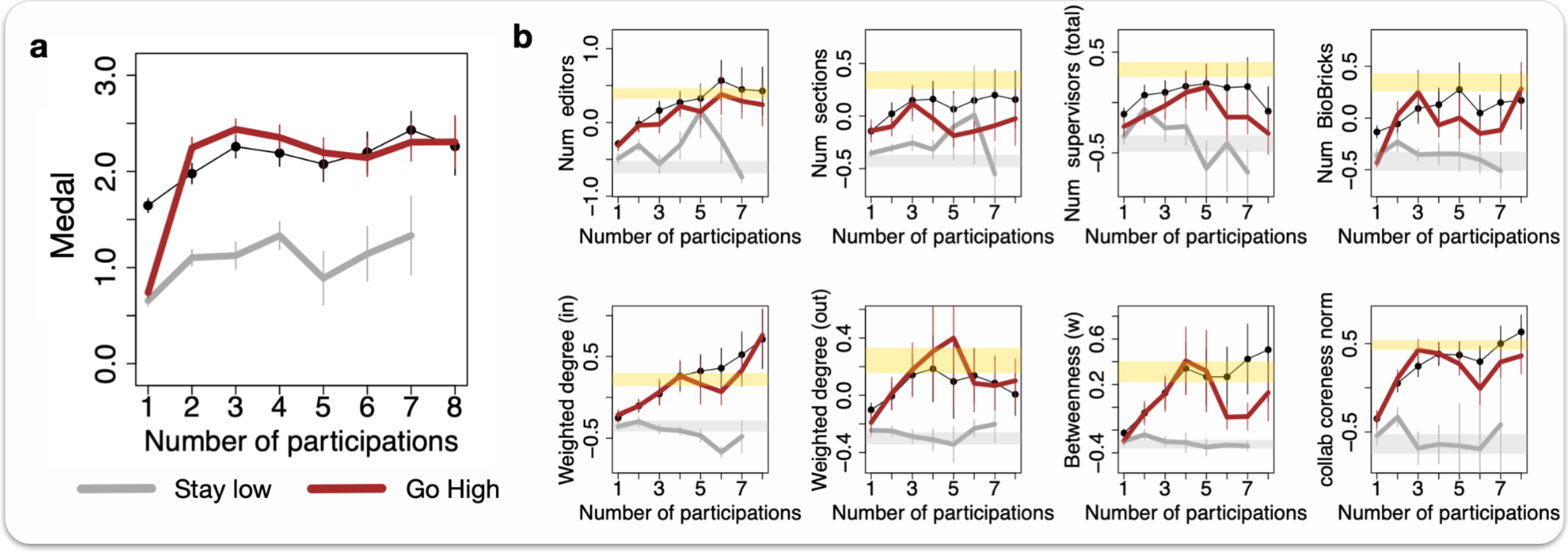

## Figure 7. Recovery from an early setback

**a.** Team success (medal) as a function of number of participations (black line, average value and standard error), as Fig 6b. We showcase two subgroups of teams with an initial setback (no medal or bronze medal on first participation) that eventually get a gold medal ("Go High", dark red) or never get better than a silver medal ("Stay Low", gray). **b.** Evolution of several Z-normalized team features across participations to the competition.

To understand the potential mechanisms behind the different outcomes of low initial performance teams, we investigate the change in both groups over multiple participations (Fig 7b). We find that teams that succeed after an early setback tend to exhibit the features of successful teams, a year or more before they experience success themselves, while teams that don't improve their performance seem not to learn how to organize in a way that will support improvement, and remain locked-in (Fig 7b). Therefore, while starting from similar initial conditions, teams can escape the lock-in effect by changing their behavior as early as the second participation, showing how early organizational interventions may have large and persistent effects on team success.

# Discussion

Science and innovation are increasingly team rather than individual activities. In this context, success—the production of new and useful ideas— hinges not on recruiting a lone genius but on optimizing interactions within and across team boundaries. While the shift to teams is well documented, data to understand the origins and consequences of team organization has lagged behind.

We introduced iGEM as a model system that supports rich insights into the roots of success in team science and innovation. We showed how the implementation of the competition, particularly its use of team wikis—open laboratory notebooks—provides granular documentation of team organization: static and dynamic data on internal structure, external collaborations, project choice, performance and success over many years. We highlighted how these features can be leveraged to quantify the impact of organizational decisions on team productivity and success. Furthermore, we showed that while the iGEM competition has grown in size, dynamical and structural patterns in the competition as well as the primary criteria for success have remained largely stable across time. Finally, we described how the panel structure of this dataset allows researchers to trace the long-term dynamics of individual teams and could be exploited in future work to support causal inference about the determinants of behavior and success. As an example, we showed the consequence of initial team success for long-term dynamics of improvement, highlighting the importance of facilitating the onboarding of new teams into the larger network to create the conditions for recovering from an early failure[59].

The relative stability of the competition and the panel structure of the data can be useful for future work interested in the organizational roots of innovation. For instance, the large number of collaborative events recorded both within and across teams can be leveraged to constrain generative models of team organization using dynamic, multilevel, multiplex frameworks of interest to physicists, network scientists and other scholars.

Looking forward, this data can also be augmented in many useful ways to address key questions for the science of science and innovation. First, collaboration in science is typically studied within teams, despite pervasive interactions across their boundaries as research groups reach out to peers for help[60–64]. These cross-team interactions are well documented in iGEM and accompanied by text that allows the network to be filtered to study particular types of interaction, or to answer questions about the interactional content of social interactions[65]. Second, beginning in 2015, a change in medal requirements significantly increased the density of the iGEM inter-team network, allowing for the causal investigation of how (typically hard to find) exogenous changes in whole network structure affect team behavior and success[66,67].

While we present data related to team performance via medals, in 2014 a sophisticated judging rubric was introduced that comprises more than 60 criteria that can offer even more fine-grained insights into specific dimensions of scientific performance evaluation, including creativity and impact. Contributions to the Biobrick repository as well as publication, patents and start-up formation data related to team projects can also enable quantifying the novelty and long-term impact of team projects and the overall team-of-teams behavior of the competition. The trajectory of competition alumni relative to other peers in the field can also be used to assess its impact on industry practices[68]. Finally, the competition has also developed an increasingly sophisticated system of requirements and rewards related to social responsibility that could offer insights into scientific governance.

Altogether, the iGEM competition offers a window into the organizational roots of innovation by providing an unprecedented testbed to analyze team processes and performance in an open science and engineering context**.** With a growing body of teams participating, future competitions will allow for further qualitative and quantitative analyses to complement the publicly available open lab wiki notebook data[69,70]. This suggests opportunities to extract even richer insights towards data-informed strategies for building interdisciplinary open innovation ecosystems.

# Methods

## Data collection

The dataset was collected through the creation of scripts to automate the extraction of content from the official iGEM websites and wikis. Here we explain the underlying page structures and scripts that allowed us to generate this dataset. User names have been anonymized by hashing their user names using the MD5 hexdigest function with a salt, in accordance with the ethical guidelines provided by the IRB number 170602.

### Team metadata

When a team registers to the iGEM competition, it must provide information about its composition (list of team members and their roles in the team, see section "User Metadata"), location (Country, Continent), Division (e.g. Overgrad, Undergrad, Masters, High School), and application track. This information is made publically available on the iGEM website. We built a custom crawler and parser in python relying on the built-in HTMLParser and BeautifulSoup 3 libraries to automate the extraction process by targeting the iGEM Common Gateway Interface page https://igem.org/Team.cgi. We note that there are a total of 4 teams for which the Division is declared as "Unspecified" in 2014 and 2015 in the iGEM database (Fig 2b). This yielded 2,406 teams across years 2004-2018.

## User metadata

During the registration process, each iGEM team member is associated with one of the following roles: 'Student Members', 'Student Leaders', 'Instructors', 'Advisors', 'Primary PI', 'Secondary PI', and 'Other' (Fig 2c). We also found wiki editors who were not declared in the official team roster, and annotated them as 'Unspecified' members. The team member roles cover 40,520 role assignments across 30,972 unique participants.

## Team results

The iGEM competition awards teams with medals, prizes, and designates for each Division finalists and a winner. These judging results are readily available on the iGEM website. We built a custom parser to extract medals, awards, finalists and winners for each competition cycle. In the manuscript when referring to the medal obtained we used an ordinal metric with values 0 for no medal, 1 for bronze, 2 for silver, and 3 for gold. Not all divisions have medals, and some divisions have medals some years but not others. When a division has no medals, the dataset indicates it as “blocked”. The “high school” division had no medals awarded until 2015, after which medals were awarded. For track awards, we classified them as four types: “first”, when a team wins the award, “second”, when the team is a runner up, “regional”, when the award is delivered to a regional championship and “other”, for nominations and advancement to a regional championship. To note, in 2011 iGEM only delivered regional awards. When considering significant awards in the main text, we focus on award winners (classified as “first” or as “regional” for 2011). The dataset comprises a total of 1,630 awards documented.

## BioBrick information

We extracted the BioBricks created by each team from the iGEM BioBrick database. To do so we utilized the iGEM provided CGI page located at http://igem.org/Team_Parts.cgi and constructed a custom parser with the default python library HTMLParser. For each team we report the IDs of the BioBricks that they produced, with a total of 33,687 BioBricks.

## Team wiki data

One of the requirements of the iGEM competition is that each team writes a collaborative and openly accessible laboratory notebook. This notebook must contain the necessary information to reproduce the results obtained by the team during the competition. For each year of the competition, iGEM provides and hosts a MediaWiki instance that teams are required to use to write their lab notebook. Due to this technological choice, the logs of all edits are stored in an information rich manner, including the timestamp, user, type, in-page location and content of the edit. In order to extract this information in an automated manner, we wrote custom python scripts that leverage the well defined HTML structure, wikimedia text formatting, as well as the API underlying MediaWiki implementations.

**Extraction of the pages structure**

In order to extract the wiki structure for a given team, we utilized the MediaWiki API endpoint Allpages (https://www.mediawiki.org/wiki/API:Allpages). We built a custom python script that automates the process of getting the data from this API endpoint for all years since 2008, when this structure was adopted by the competition. To extract the pages, we utilized the fact that iGEM enforced the following naming schema [year].igem.org/Team:[Team_name]. Of note, one team (Shenzhen SZMS, in 2013) escaped our algorithm as they did not follow that schema, and had all their pages at the iGEM root URL. The pages are used in the subsequent analyses described below. The dataset contains a total of 115,628 unique pages.

**Extraction of page edits**

In order to extract the logged event for all the users on the iGEM wiki we utilized the MediaWiki API endpoint Revisions (https://www.mediawiki.org/wiki/API:Revisions). We built a custom python script that automates the process of getting the data from this API endpoint for every page recovered from the previous step. The collected edits are classified in three actions: page creation, page edit and page deletion. Moreover, we collected the size of the edits (in bytes, a close equivalent to the number of characters changed), their sign (addition or removal of characters), as well as the User ID of the team member who performed the action. The final wiki edit data contains 2,626,485 edits from 13,655 editors. Since the iGEM HQ edits the original templates for each team wiki, edits from user "iGEM HQ" were removed prior to analyses, representing 1.7% (43,135 edits) of the original data.

**Extraction of wiki sections**

Utilizing the same API endpoint as Extraction of page edits, we can also analyze the changes in content of the page overtime. This includes the final version, as well as the progress towards this final version. Using the MediaWiki text structure we defined wiki sections as Header 2 text blocks (corresponding to == Header 2 ==). We then parsed each revision content to analyze which section was edited. In order to match each section to a unique ID, we developed a custom python pipeline which performed the following steps. First, we constructed a dictionary containing all the unique section names. Since there could be slight modifications of a section name over time, we performed a fuzzy text matching of all section names using the FuzzyWuzzy library and a Levenshtein distance cutoff of 0.80. If two names matched, we grouped them together. If the names did not match, we checked if the content of two sections was similar using a fuzzy text matching between the section contents with a cutoff of 0.80. If the content passed this test, we grouped the sections together. Finally, we assigned a unique ID to each group, and used this ID for the revision edits. This process allows our algorithm to follow the edition of a page by multiple users through time at the section level, despite potential changes to the text and section title, up to our cutoffs levels. Our algorithm extracted 196,252 sections, with an average of 22 sections per page.

**Extraction of the team mentions**

Teams mention other teams in their wiki project page to refer to their interactions, such as exchanging materials and advice, or joint-problem solving and socializing. To extract these mentions, we first collected the entire text from each team project wiki. iGEM maintains a public webpage listing for each team the links to all pages of their wiki website in the form [year].igem.org/wiki/index.php?title=Special%3APrefixIndex&prefix=Team%3A[Team_name]. We used the *Extractor* function from the *boilerpipe* Python library to extract the html from all these pages as content using the *KeepEverythingExtractor* option. We then used the *justext* function from the *justext* Python library to remove html tags. Finally, we used the English pickle tokenizer ('tokenizers/punkt/english.pickle') from the *nltk* library to extract a list of sentences.

We then looped over the extracted sentences to collect team mentions while accounting for variation in spelling. For each sentence, we searched for mentions of iGEM team names from the same competition using fuzzy matching. We used the *partial_ratio* function of the fuzzy wuzzy Python library for substring matching and set the Levenshtein distance cutoff at 0.90. However, when the sentence is a page or section title, the length of the sentence might be smaller than the length of the team name we are comparing with. In such edge cases, the partial ratio generates false positives by matching teams to common title substrings, like the title Software matching to all software track teams (USTC_Software, Michigan_Software, SYSU-Software etc.). We instead use regular fuzzy matching with a relaxed cutoff of 0.70. Finally, for each fuzzy match we reported the source team name, the target team name, the name of the wiki page where the sentence is present, and the year.

Some pages in the wiki page list are residuals of template pages. iGEM creates these pages for each team and lists examples of teams from previous years as a reference. However, some teams do not edit these template pages, but simply do not provide links to them within their wiki website, making them invisible to the public eye (but not to our method). To remove such cases, we first collected instances where a target team is mentioned by different source teams using the exact same sentence. If these instances occurred more than 40 times a given year, we removed this match from the list. In addition, we used contextual information from template pages to prune out potential false positives. The template pages refer to previous successful teams by mentioning the team name and the prior year in which they participated in the form [Year] [Team name][2]. We used a regular expression to identify the sentences from the unedited template pages containing a year and a team name, and removed the corresponding instances if they appeared more than 10 times that given year. Finally, we also removed sentences which correspond to open source license statements as they reference MIT, which is also an iGEM team, using matches to the following phrases: "Licensed under MIT", "Licensed under the MIT", "MIT License", "MIT license", "COPYRIGHT OWNER OR CONTRIBUTORS BE LIABLE", "PROVIDED BY THE COPYRIGHT HOLDER", "(Custom Build)", "WITHOUT WARRANTY OF ANY KIND", "Licensed MIT", "MIT/GPLv2 Lic", "MIT/GPL2", "Released under the MIT", "Dual licensed under", "License: MIT", "MIT Press", "MIT_License", "LICENSE-MIT", "MIT/BSD license", "MIT @license", "MIT Technology Review", "THIS CODE IS PROVIDED ON AN *AS IS* BASIS", "MIT (c)".

[2] See for example https://2018.igem.org/Team:Aachen/Applied_Design

To generate the edge weights of the inter-team mention network, we counted the number of unique sentences where a source team has mentioned a target team on their wiki. The resulting network consists of 15,423 collaborations.

# Manual curation

## Performance of teams in 2004-2007

While the standard wiki structure that we describe above only allows us to automate data extraction since 2008, earlier versions of the iGEM website provide medal information for 2007 and significant awards for 2006-2007. This performance data was manually curated from the relevant iGEM pages for the corresponding years and added to the final dataset.

## Dataset corrections

Manual verification of the dataset during the validation of the data exhibited 19 teams where our parser failed to extract data. Due to the unavailability of the API since 2019, we could not extract the wiki sectionID for those teams. First, five high school teams from 2013 had an Undergraduate URL in the iGEM database, resulting in a failed attempt at parsing their wiki. We manually corrected for the error, ran the crawler and injected that data back into the wiki edits table. Moreover, one high school team from 2013 (Shenzhen SZMS) was registered at the root of the wiki namespace, making it impossible to discern which pages are from their team, and which are from iGEM HQ. Second, in 2012 iGEM added a special entrepreneur track, under the subdomain 2012e.igem.org. Our original crawler did not take this into account, therefore we corrected it and then added the wiki edits for those teams.

Finally, we noticed that three teams from 2012 (Peru_BioE, sentegen, UANLe_Mexico) were annotated as “Accepted” in the iGEM website, yet they seem to have withdrawn (team roster page specifying that the registration was incomplete, and team wiki page not edited from the template). We therefore removed them from our dataset.

# Technical Validation of the dataset

Our pipeline was built with error catching checkpoints allowing the code to retry any failed access, as well as outputting any unresolvable error. Moreover, the pipeline checks that the total number of parsed teams equals the number of registered teams on the iGEM website. In order to verify that our automated extraction pipeline functions as intended, we compared the teams extracted by our pipeline from the iGEM yearly mediawiki instances (Wiki data), as shown in Figure S4. We observed an exact match for all years except for the Shenzhen SZMS team in 2013 (see “Dataset correction” section of the Methods).

# Analysis

## Adjusting for scale variance across time

In order to take into account the variation of average team features across competition cycles, we define for each feature $f$ of a team $i$ a Z-score comparing it to all teams from the same year $y$:

$$Z_{f,i} = \frac{f_i - \mu_{y,f}}{\sigma_{y,f}}$$

where $\mu_{y,f}$ is the average feature value for year $y$ and $\sigma_{y,f}$ its standard deviation. We then use this Z score for the descriptive statistics of team improvement across competitions (Fig 6-7).

## Measures of centrality in the inter-team mention network

We define for each competition year the inter-team mention network as a weighted, directed network where edge weights correspond to the number of mentions of a given team to another team in their wiki. Network centralities are evaluated using the igraph R library. The weighted degree is computed with the “strength” function, with the mode parameter set as “in” for in-coming links and “out” for outgoing links. In the manuscript we consider a normalized version of this quantity, where the strength is divided by the average strength the same year (see Figure 6e). Weighted clustering is computed using the “transitivity” function. We also consider two measures of global centrality, namely coreness and betweenness centrality. Coreness is a measure of the embeddedness of a node within the network, using a “russian doll” approach. The coreness of a node is k if it belongs to the k-core, i.e. the maximal subgraph in which each vertex has at least degree k, but not to the (k+1)-core. We used the “coreness” function and normalized the obtained value to 1 by dividing by the maximum value each year. Betweenness centrality quantifies the number of shortest paths going through a team. In our case we use weighted shortest paths, with weights equal to the inverse of the number of mentions so that they are interpreted as distances.

### Team multiple participation

Much like other teams in areas like sports, individual players may rotate, but when the host organization (which is often coupled to the coach and other institutional knowledge) stays the same, they are considered the same "team". As such, in order to track team progress throughout multiple participations, we collect for each team all the years for which we find their team name in the competition roster. For each team, we sort their participations by chronological order, and collect for each feature the yearly corrected Z-scores across these participations. We finally average these values over all teams with a given number of participations $p$ to the competition to get the normalized trend.

When considering recovery from initial failure, we restrict the study to teams that have participated at least 2 times in the competition.

## Code availability

Due to the sensitive information contained in gathering non anonymized datasets of human data, the authors cannot provide the code used to generate the dataset used in this study, in accordance with the ethical guidelines provided by the IRB number 170602. However, if a laboratory is interested in utilizing this code and can provide the necessary documentation on human studies, the authors will share access to the necessary scripts.

## Data availability

The data is available on Zenodo at https://doi.org/10.5281/zenodo.10047683 as an ensemble of CSV files, with a README.MD file which describes each table, their columns names and data types.

# Acknowledgements

We would like to thank past and current members of the iGEM Foundation staff, in particular Kim De Mora, Anita Sifuentes, Meagan Lizarazo, and Randy Rettberg, for their help in understanding the data, as well as providing guidance to use their API endpoints efficiently. We also thank Peter Carr for his careful reading of the manuscript and constructive comments alongside other members of iGEM's volunteer committees and participants who have provided feedback at various stages of this project. We thank Dashun Wang and Brian Uzzi for insightful comments on the study. We thank Thomas Landrain for his key contribution to the thought process that led to the study of the iGEM competition, and for helping coordinate contacts between the authors and the iGEM headquarters. Thanks to the Bettencourt Schueller Foundation long term partnership, this work was partly supported by the LPI Research Fellowship to Marc Santolini.

Megan Palmer also received support from Open Philanthropy that enabled her and members of her research team to help conduct this research. This work was also supported by the French Agence Nationale de la Recherche (ANR), under grant agreement ANR-21-CE38-0002.

# Supplementary Information

## Additional details on the iGEM competition

In this section we provide further details on the iGEM competition structure.

The basic challenge of the iGEM competition is for teams to design and build synthetic biology-based solutions to real-world problems.

Teams are typically hosted by a larger institution —a high school, university or community lab— and are separated into divisions based on whether they are composed of high school, undergraduate or overgraduate members. Each team has a principal investigator (PI), which for university teams is often a faculty member, along with other instructors and advisors that provide guidance without directly performing work on the team's project.

The competition runs on a similar cycle each year. Teams assemble, begin to acquire funding and choose problems to work on prior to registering for the competition in April and May. The competition begins officially in May when a standard set of biological components —known as BioBricks— are distributed. Team work takes place through late October culminating in a "Jamboree" where teams present the fruits of their labor to judges and peers, and receive awards.

Teams are evaluated on a common rubric which dictates the outputs expected of all teams and therefore a core set of tasks they need to perform to succeed in the competition. But teams also have discretion in generating and selecting projects. Different choices lead to different application areas and task sets. One team may choose to develop a biology-based diagnostic test while another engineers microbes to decompose plastic waste.

These choices are organized by competition tracks (e.g. diagnostics or remediation) that allow teams to be compared within and across application areas[31].

In order to obtain awards, teams must produce and document the function of a BioBrick part that is added to a growing repository available to teams in subsequent years. Regardless of the application area they choose, teams need to design a novel project, develop it to a minimum level of functionality and present evidence supporting their claims. Teams also need to perform a range of other tasks such as satisfying safety standards, considering use cases, and integrating those considerations into the design of their projects. Finally, in addition to a presentation and poster, teams must document the process and results of their work in a digital laboratory notebook hosted on a publicly available wiki website.

# Performance Evaluation

Performance is evaluated by panels of judges based on written documentation, a poster and an oral presentation according to a common and publicly available rubric. While the core criteria have remained constant, the rubric has become more detailed over time with around 60 criteria in recent years. These criteria are communicated to teams on the official iGEM website and through other documentation. To apply these criteria, iGEM selects experts from across academia, industry, government and other settings to judge the projects. Judges form panels of six at random, conditional on a need to avoid conflicts of interest and have balanced and relevant expertise to evaluate each team. Further details can be found in the Performance Evaluation Section.

***Performance Levels:*** Success in the competition depends on judge evaluations and is recognized through multiple rewards. While there have been modifications over the lifetime of the competition the core criteria have remained mostly consistent (Fig S2-3). Additional details are provided in Table S1.

- Medals (bronze, silver, and gold) that are awarded non-competitively for demonstrating pre-specified achievements. There are no limits to the number of medals that can be awarded.
- Prizes that are awarded competitively for demonstrating excellence in a number of areas. There are two general types of prizes: (i) track prizes for best project in a team-specified application area "track" (ii) special prizes for best-in-class performance in certain aspects of the competition, such as best presentation, or best “BioBrick” part. Since 2015 nominees for awards have also been announced, which we consider as prizes in their own right. Teams are divided into highschool, undergraduate and overgraduate fields and compete only against teams in their field.

- “Grand prize” winners that are awarded competitively across all teams and tracks. Undergraduate teams compete for Winner, 1st Runner Up, 2nd Runner Up. Overgrad teams compete for Winner and 1st Runner Up. High School teams compete only for the Winner.

***Judging:*** The judging process has also evolved to become standardized as the competition has grown. It has consistently involved panels of approximately 6 volunteer experts across academia, industry, government and civil society around the world. While judging occurred by consensus judging among expert panels in early years, as the competition has grown, judging moved to a standardized blind online voting system. There are some aspects of the judging process that are proprietary, including the weighting and final analysis of scores, but in general the following process is followed:

- Judges each assign a medal recommendation and a final score is computed across all judges to assign a final medal to the team.
- For track, special and grand prizes, judges each evaluate teams across several project criteria, and a final score is computed across all judges.

| Performance Level | 2018 requirements for standard tracks | Evaluation Method |
| --- | --- | --- |
| **No Medal (Participant)** | • Register for the competition | N/A |
| **Bronze** | • Register for the competition<br>• Attend Jamboree<br>• Develop a wiki<br>• Produce a poster<br>• Give a presentation<br>• Complete Judging Form<br>• Attribute work<br>• Complete interlab measurement study OR add new characterization data for an existing BioBrick part | • Non-competitive<br>• Requires all criteria met |
| **Silver** | • Demonstrate new BioBrick part(s) "work(s)"<br>• Collaborate with a team<br>• Demonstrate careful and creative thinking about whether project is responsible & good for the world | • Non-competitive<br>• Requires all criteria met |
| **Gold** | Two of the four:<br>• Integrate thinking about whether project is responsible & good for the world into project purpose, design, and/or execution<br>• Create a new BioBrick part that is an improvement on an existing BioBrick part<br>• Model your project<br>• Demonstrate that engineered system works | • Non-competitive<br>• Requires a subset of criteria met (3 out of 4) |
| **Prizes: *Best in a "Track"*** | • Diagnostics<br>• Energy<br>• Environment<br>• Food & Nutrition<br>• Foundational Advance<br>• Information Processing<br>• Manufacturing<br>• New Application<br>• Therapeutics<br>• Open (Special Track)<br>• Software (Special Track) | • Competitive<br>• Projects judged on about 10 criteria |
| **Prizes:** Special | • Best Education & Public Engagement | • Competitive |
| Prizes for “Best in Class” in an area | ● Best Hardware<br>● Best Integrated Human Practices<br>● Best Measurement<br>● Best Model<br>● Best New Basic Part<br>● Best New Composite Part<br>● Best Part Collection<br>● Best Plant Synthetic Biology<br>● Best Poster<br>● Best Presentation<br>● Best Product Design<br>● Best Software Tool<br>● Best Supporting Entrepreneurship<br>● Best Wiki | ● Projects judged on about 5 criteria |
| **Grand Prize “Winner”** | ● Grand Prize Undergraduate<br>● First Runner-Up Undergraduate<br>● Second Runner-Up Undergraduate<br>● Grand Prize Overgraduate<br>● First Runner-Up Overgraduate<br>● Grand Prize High School | ● Competitive<br>● Projects judged on about 10 criteria |

# Team size

Teams are composed of student members (the active members who conduct the project) and more senior instructors and Principal Investigators that guide and mentor the team. Team size can therefore refer to the total number of registered team members ($N_{registered}$, including students and supervisors), the number of registered students ($N_{students}$, excluding supervisors) or the number of students editing the wiki ($N_{editors}$). We find that the number of wiki editors scales with the larger team size, with $N_{editors} \sim N_{students}^{0.53}$ and $N_{editors} \sim N_{registered}^{0.61}$ (Fig S6). But are the editors “ghost writers” for the rest of the team, or are they representing a bona fide core team of active team members?

To test the ghost writer effect, we compare the project output of teams for which the proportion of students editing the wiki is above or below a low threshold of 0.3. We show in Fig S9 that for both groups, measures of team performance (number of BioBricks produced, number of edits per editor) and success (number of medals and number of awards) are not significantly affected.

Finally, we compare the behavior of various performance and success measures when considering respectively $N_{editors}$, $N_{students}$, and $N_{registered}$ (Fig S10). We find that these measures display an overall greater variability as a function of $N_{editors}$ compared to $N_{students}$ and $N_{registered}$, further supporting its operationalisation as a measure of active membership in the team. This is further confirmed by the results from the modeling in Figure S19, where the number of registered team members does not affect measures of performance and success, while the number of editors does significantly so.

# Supplementary Figures

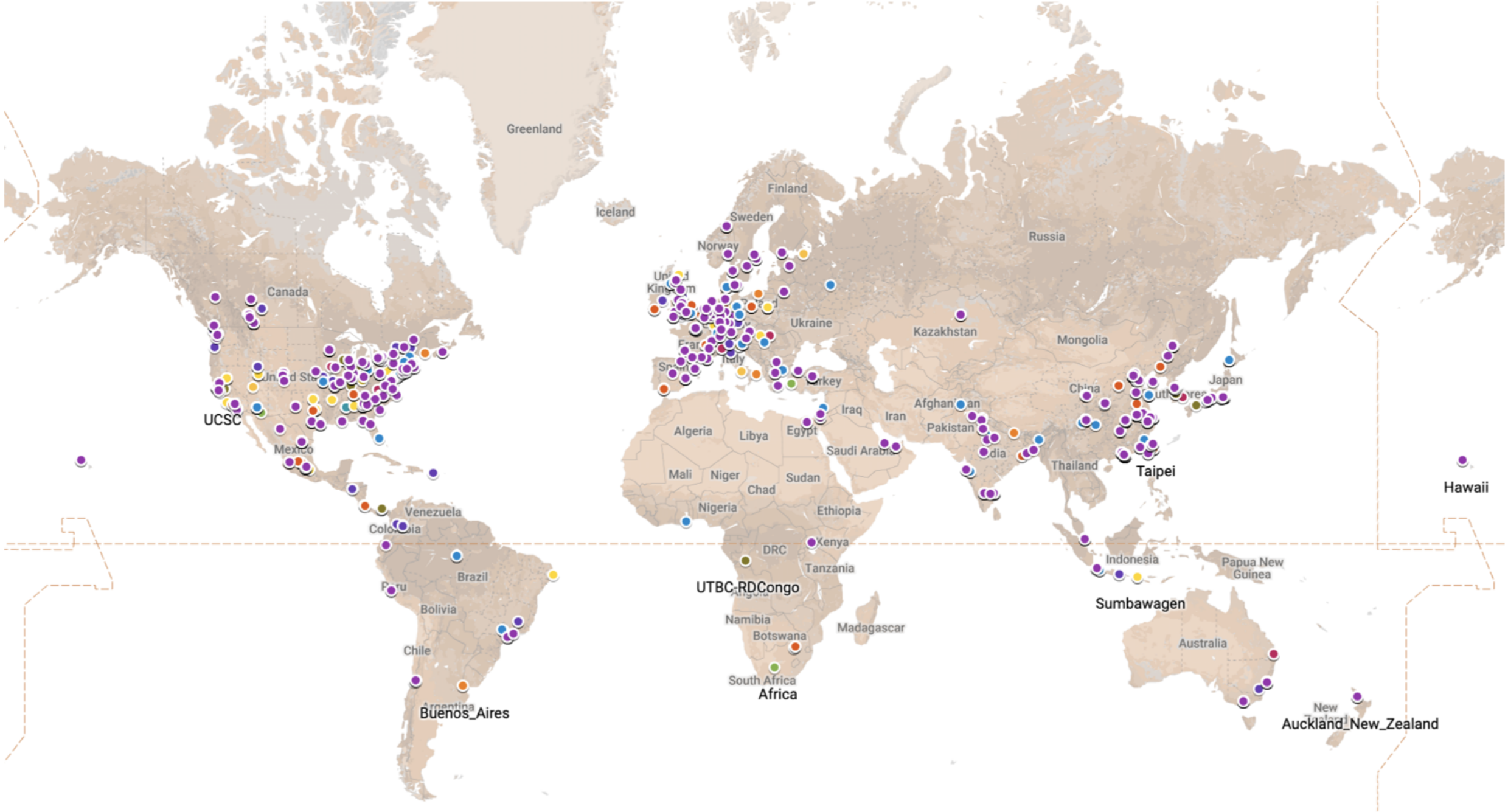

## Figure S1 - Team locations

Geographical diversity of the teams represented on a world map. Each dot corresponds to a team, and the color of the dot represents the last year the team participated in the competition.

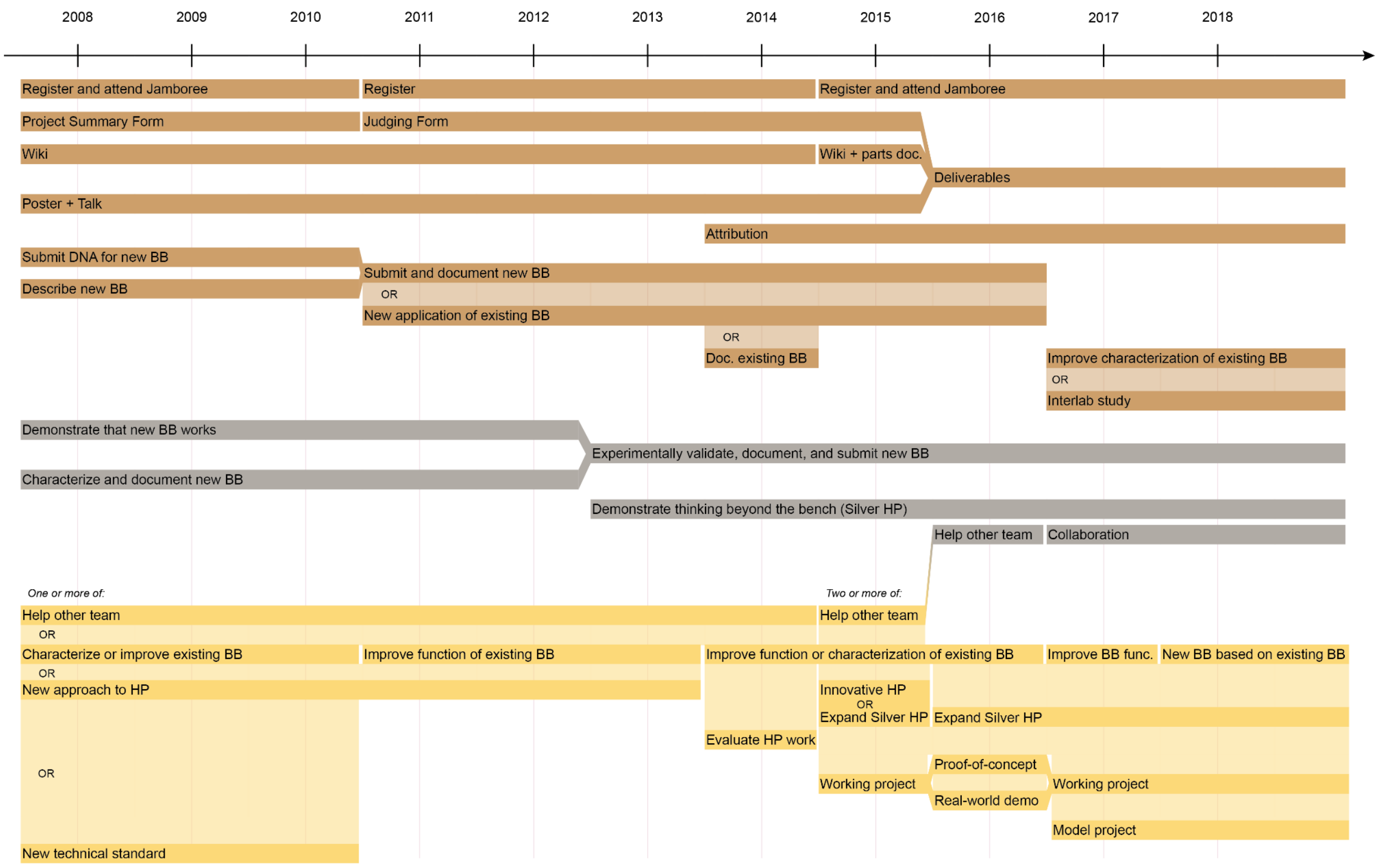

# Figure S2 - Medal criteria

Evolution of the medal criteria listed in Table S1 across competitions.

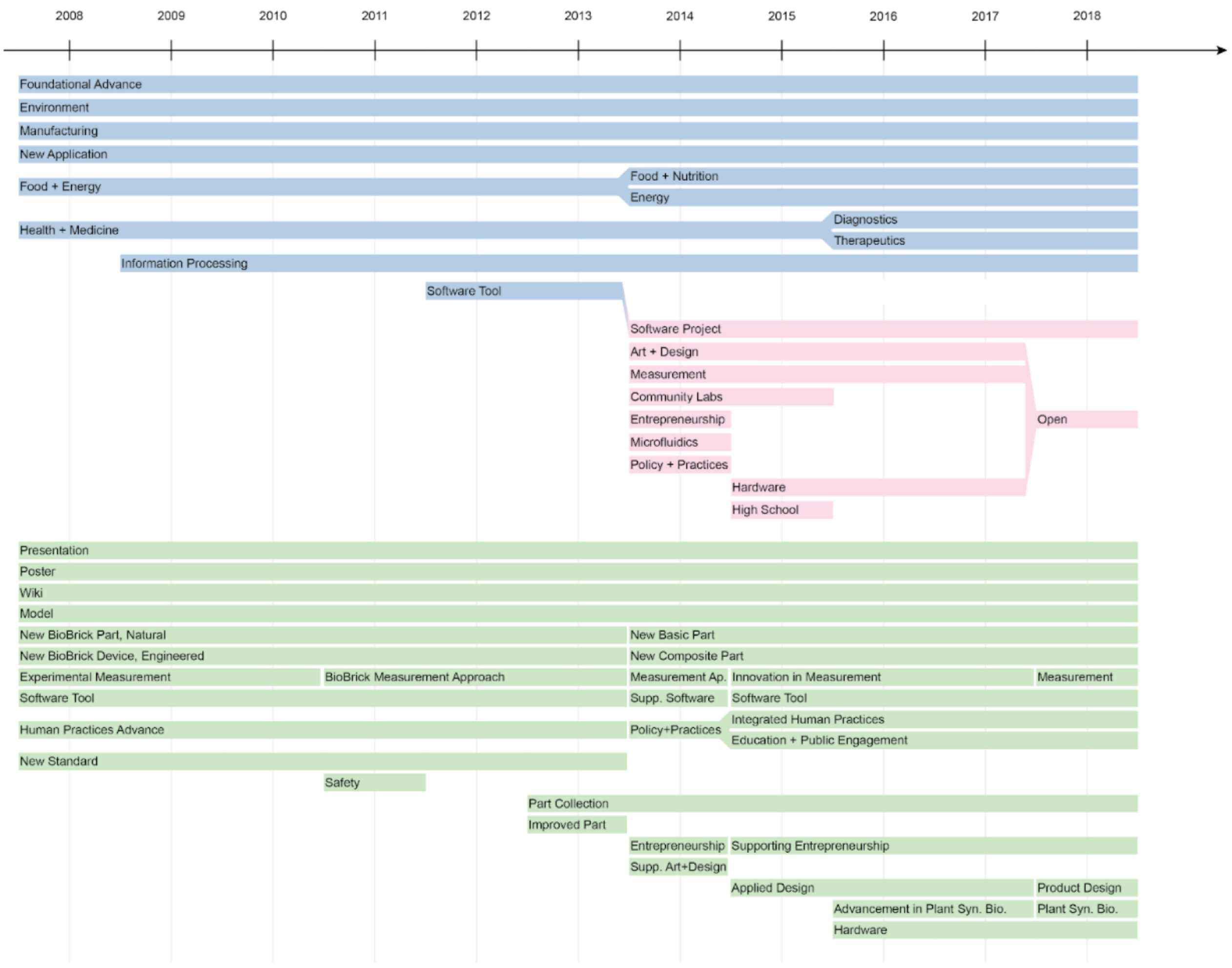

# Figure S3 - Evolution of awards

Evolution of the awards listed in Table S1 across competitions.

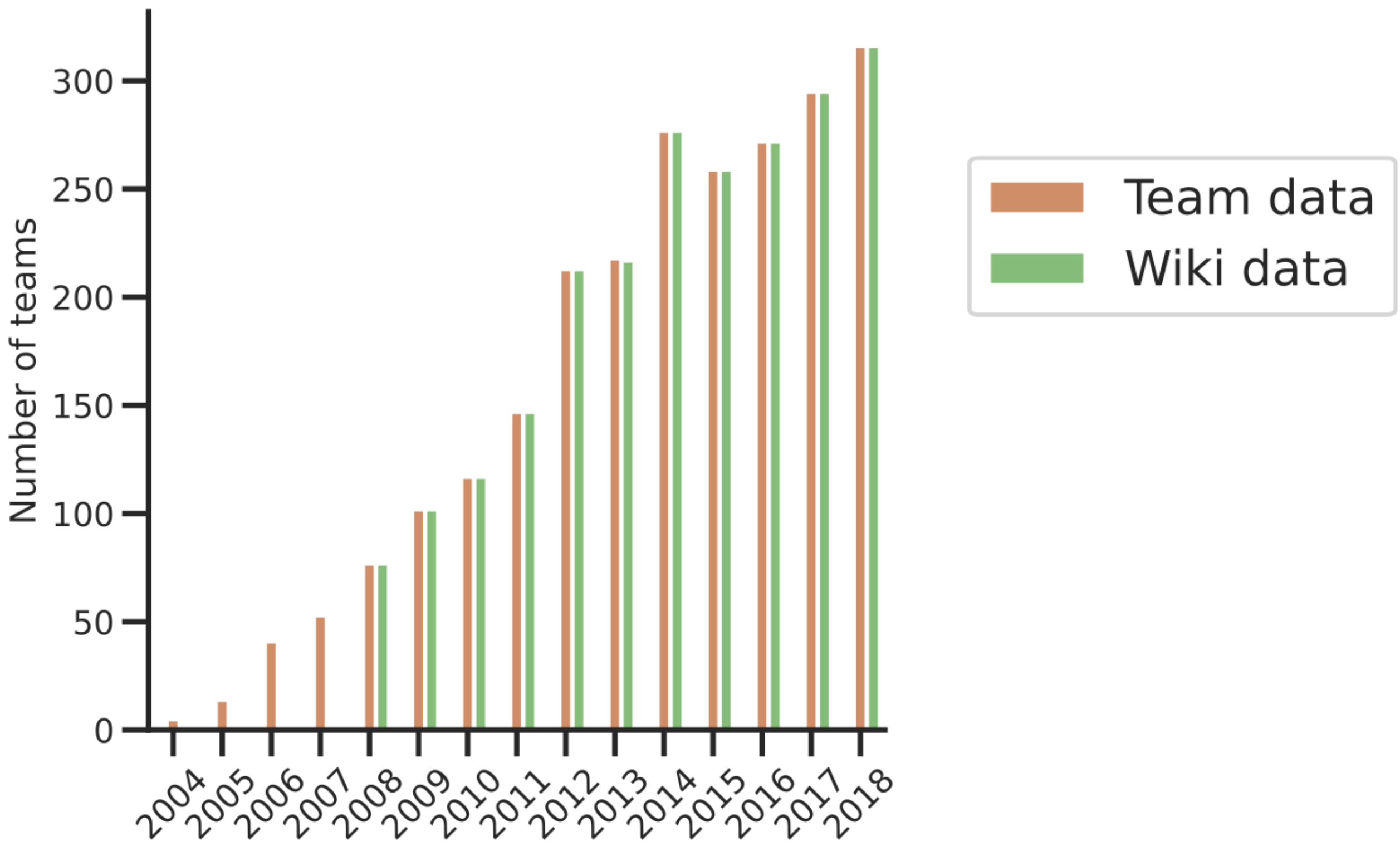

## Figure S4 - Technical validation of extracted data

Number of participating teams each year since the start of the competition. We compare the information extracted by our pipeline from the iGEM team registry (Team data) and from the iGEM yearly mediawiki instances (Wiki data).

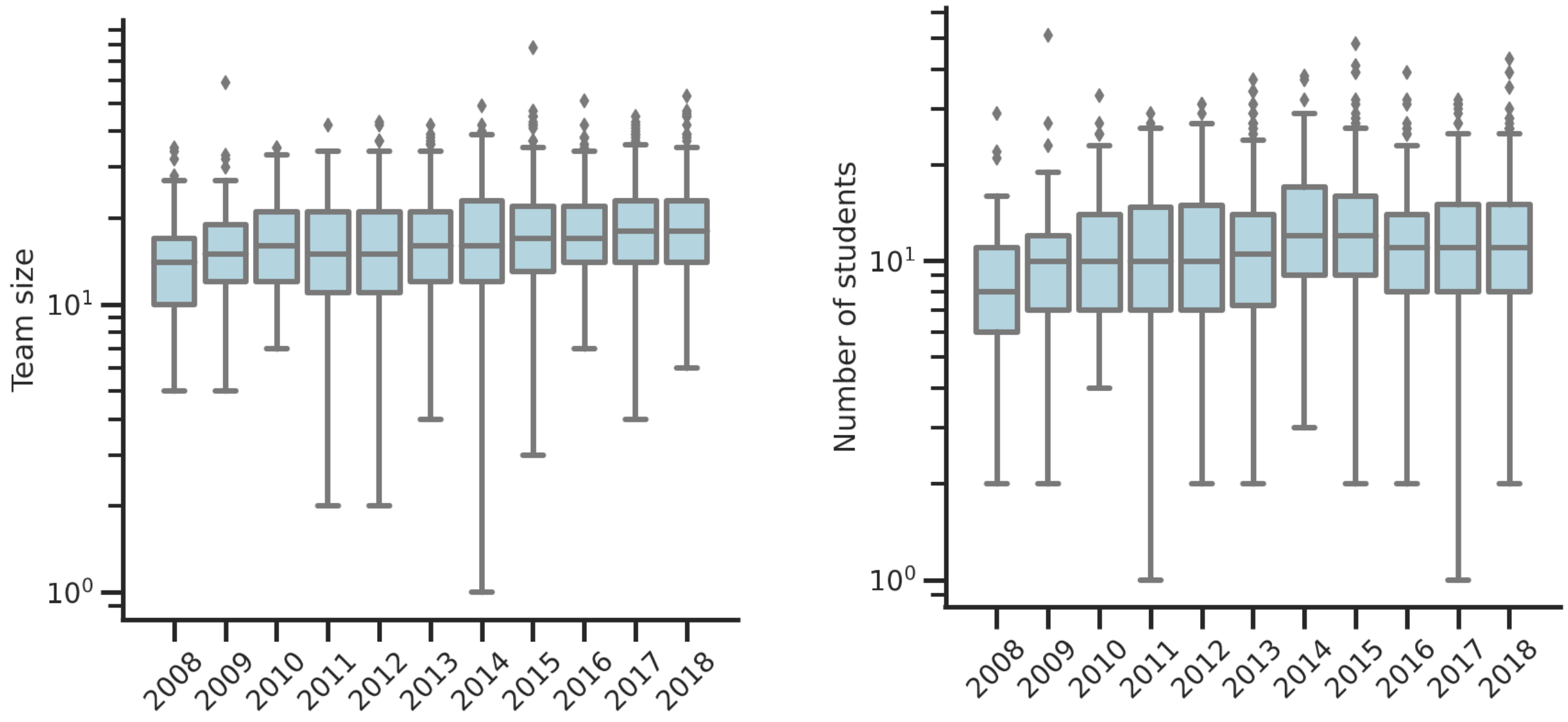

# Figure S5 - Stability of team sizes across competitions

Number of registered team members and number of team members assigned with a student role per team across competitions.

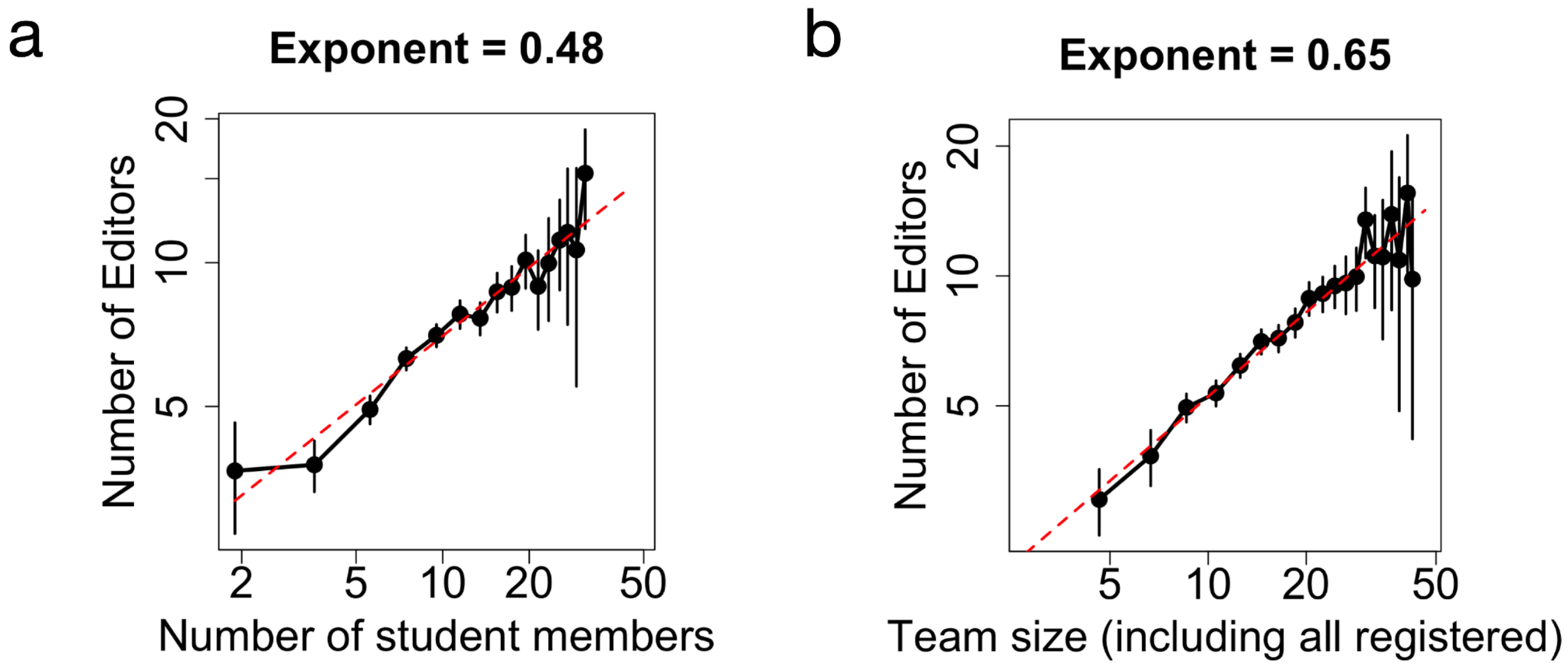

## Figure S6 - Scaling properties of the number of editors with team size

Number of editors as a function of number of student members (a) and total number of registered team members, including instructors and PIs (b). We show average and standard error across bins of size 2. We use a log-log scale to exhibit the scaling behavior. The dashed red line corresponds to a linear regression in log-log space $y \sim x^{\alpha}$, whose slope $\alpha$ is shown in the panel title.

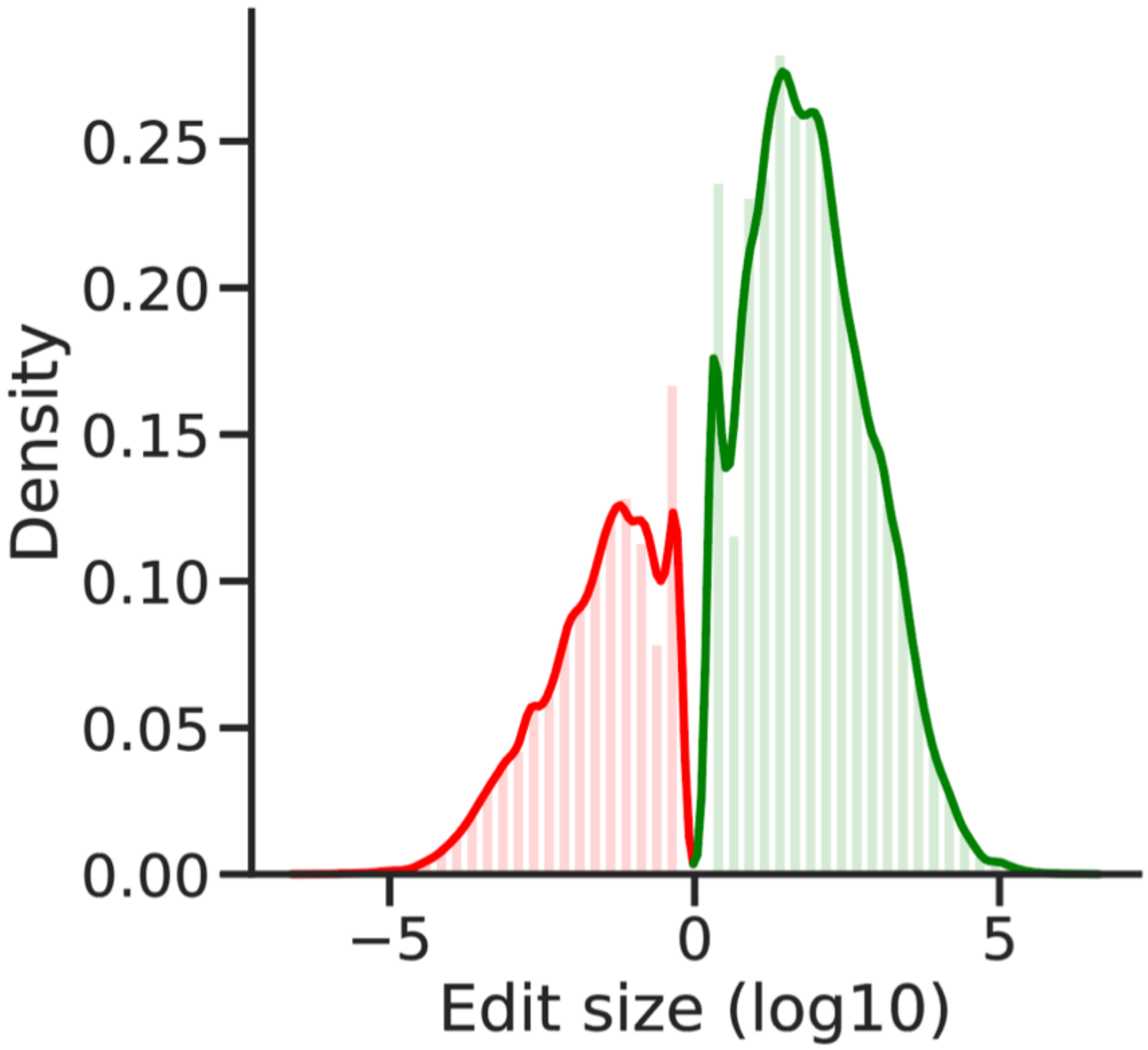

## Figure S7 - Distribution of edit sizes across teams

Distribution of edit sizes across the dataset, using a log scale. Red (respectively green) color indicates edits for which content was removed (resp. added), corresponding to a negative (resp. positive) edit size. We note that a significant proportion of edits (~19%) are of size 0, which correspond for example to minor edits (e.g. replacing one character) or the creation or deletion of a page, etc.

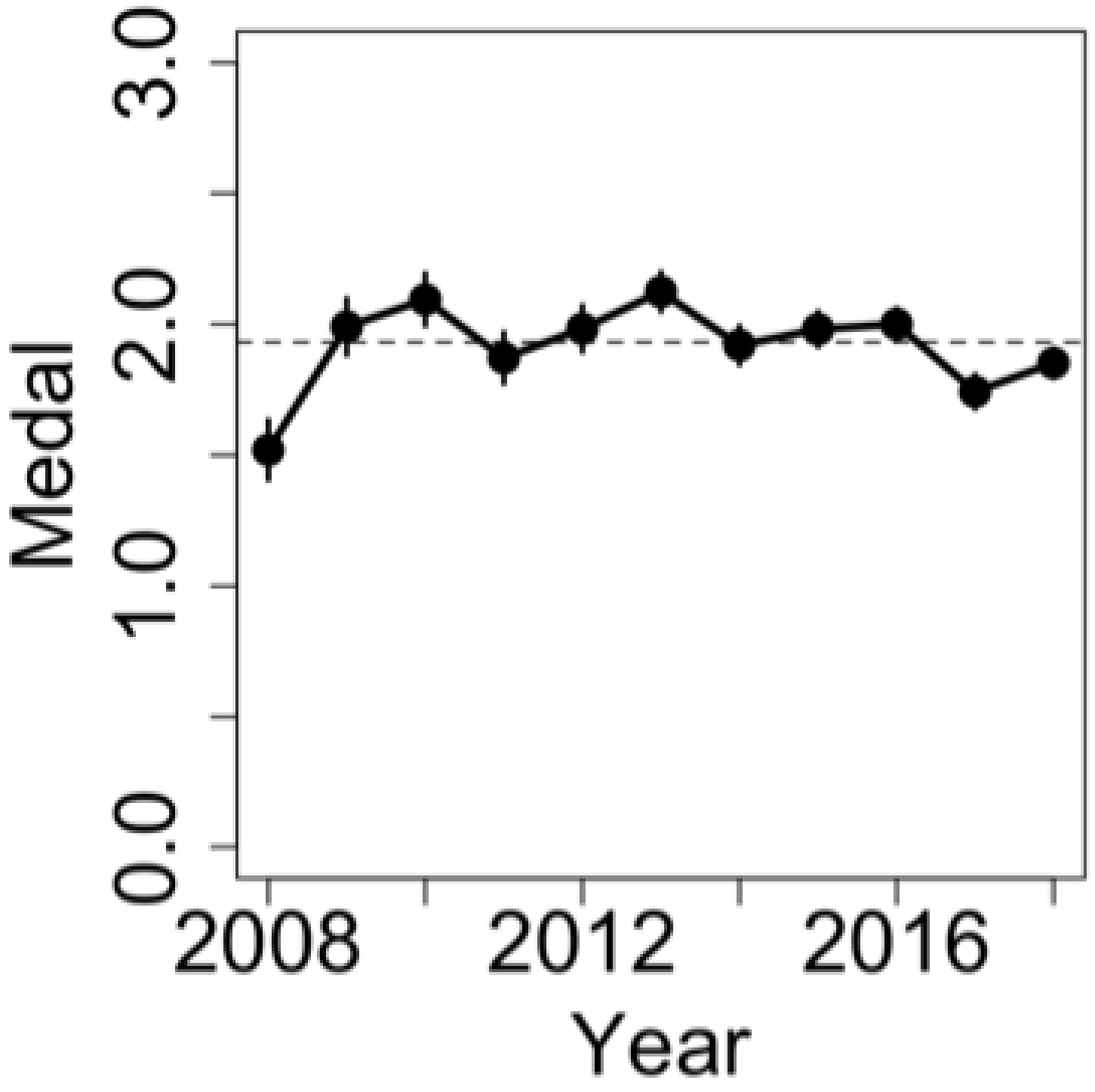

## Figure S8 - Stability of average medal awarded across years

We note that in 2008, a smaller number of gold medals were awarded, after which the typical medal quality remained stable.

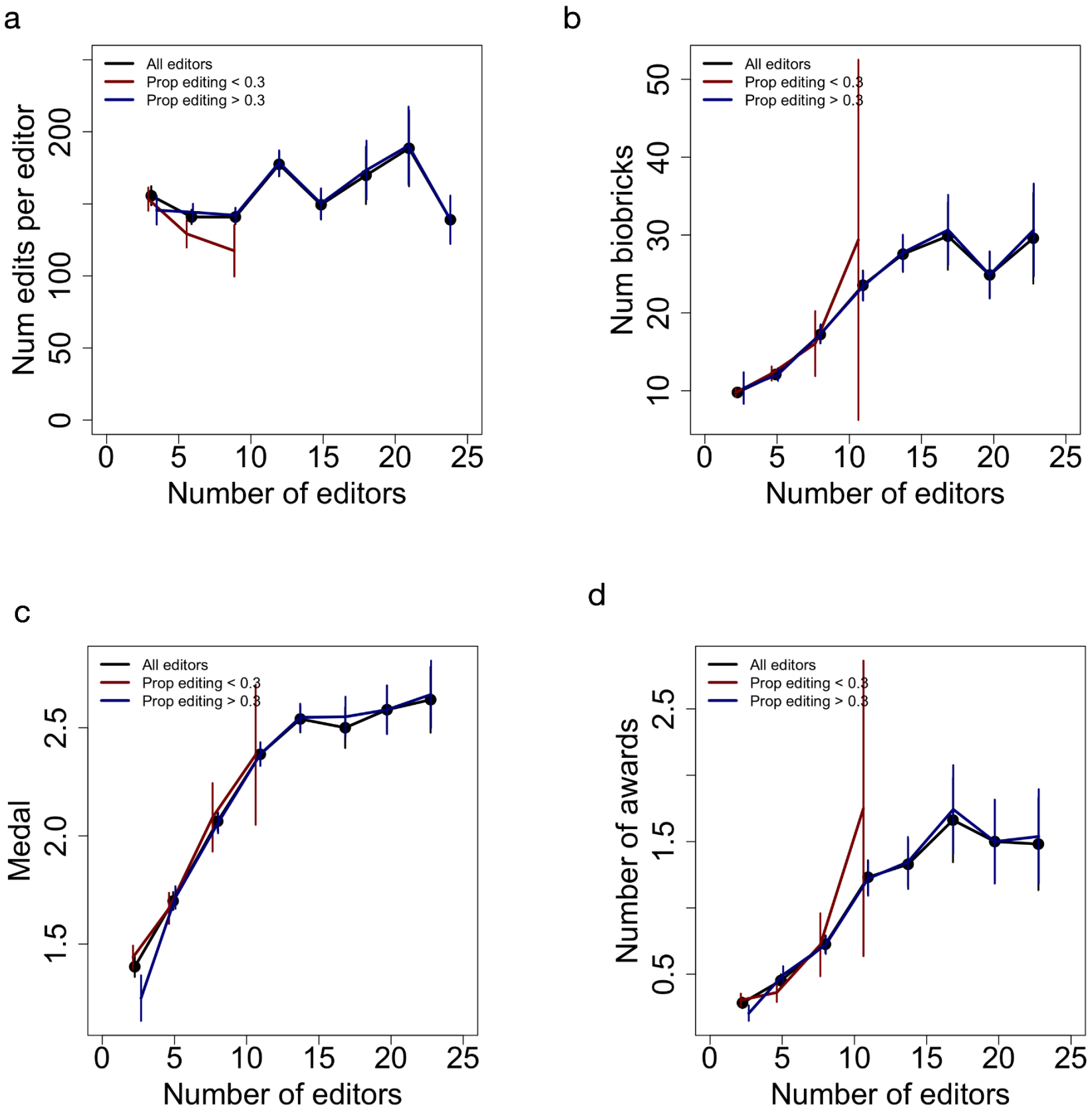

# Figure S9 - Relevance of editors as an active team subgroup

Evolution of various measures of team performance and success as a function of the number of editors. To study whether editors represent active members of the team, we distinguish teams for which the proportion of students editing is more (blue) or less (red) than a threshold of 0.3. Both cases follow the overall trend (black lines). Error bars indicate standard error.

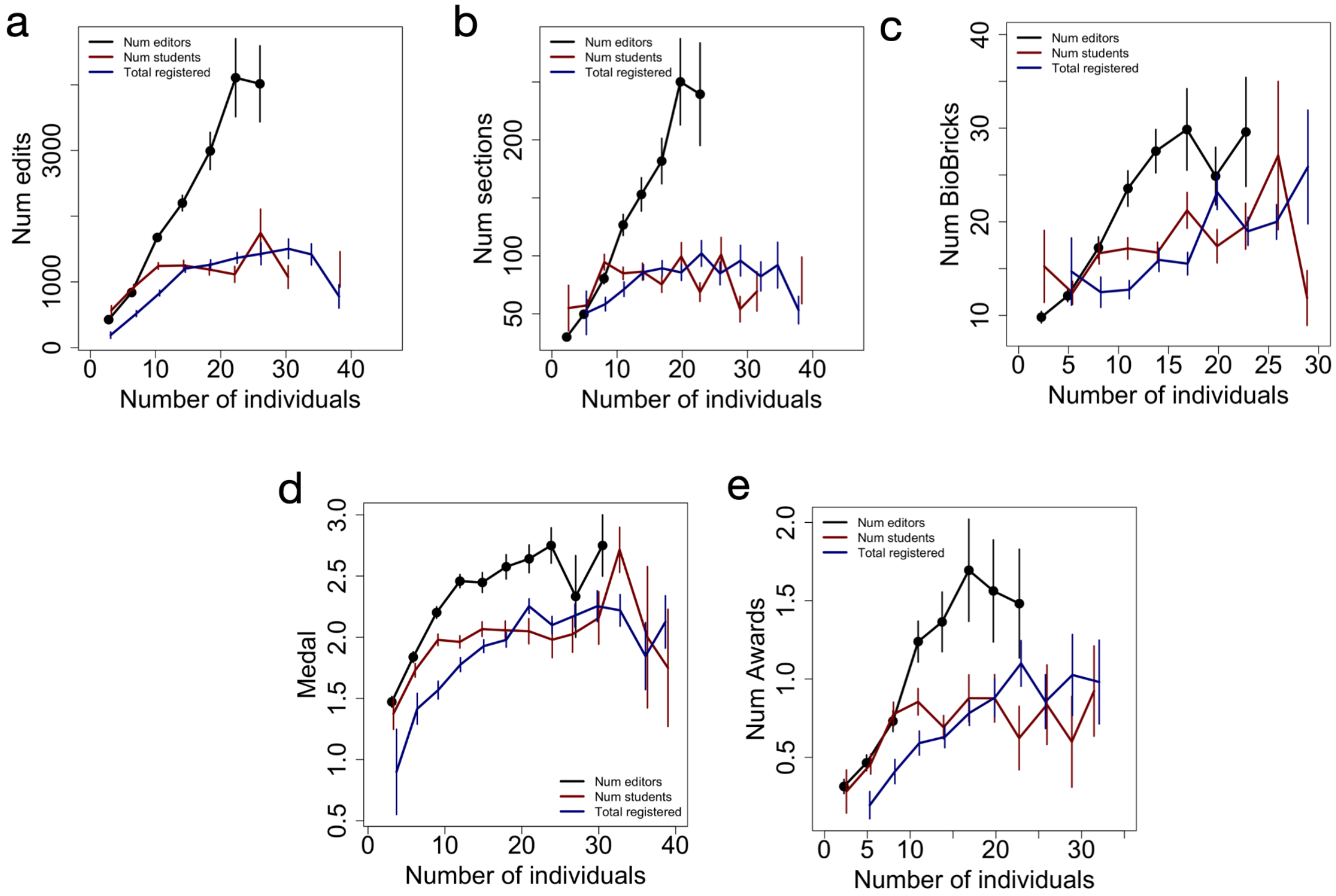

## Figure S10 - Performance and success for different measures of team size

We compare the behavior of various measures of performance, productivity and success as a function of the number of individuals measured respectively by the total number of team members registered (blue), the number of team members with a student role (red) and the number of wiki editors (black). We observe in general wider variance in the outcomes when considering editors, making them a good proxy of an active team core. Error bars denote standard error.

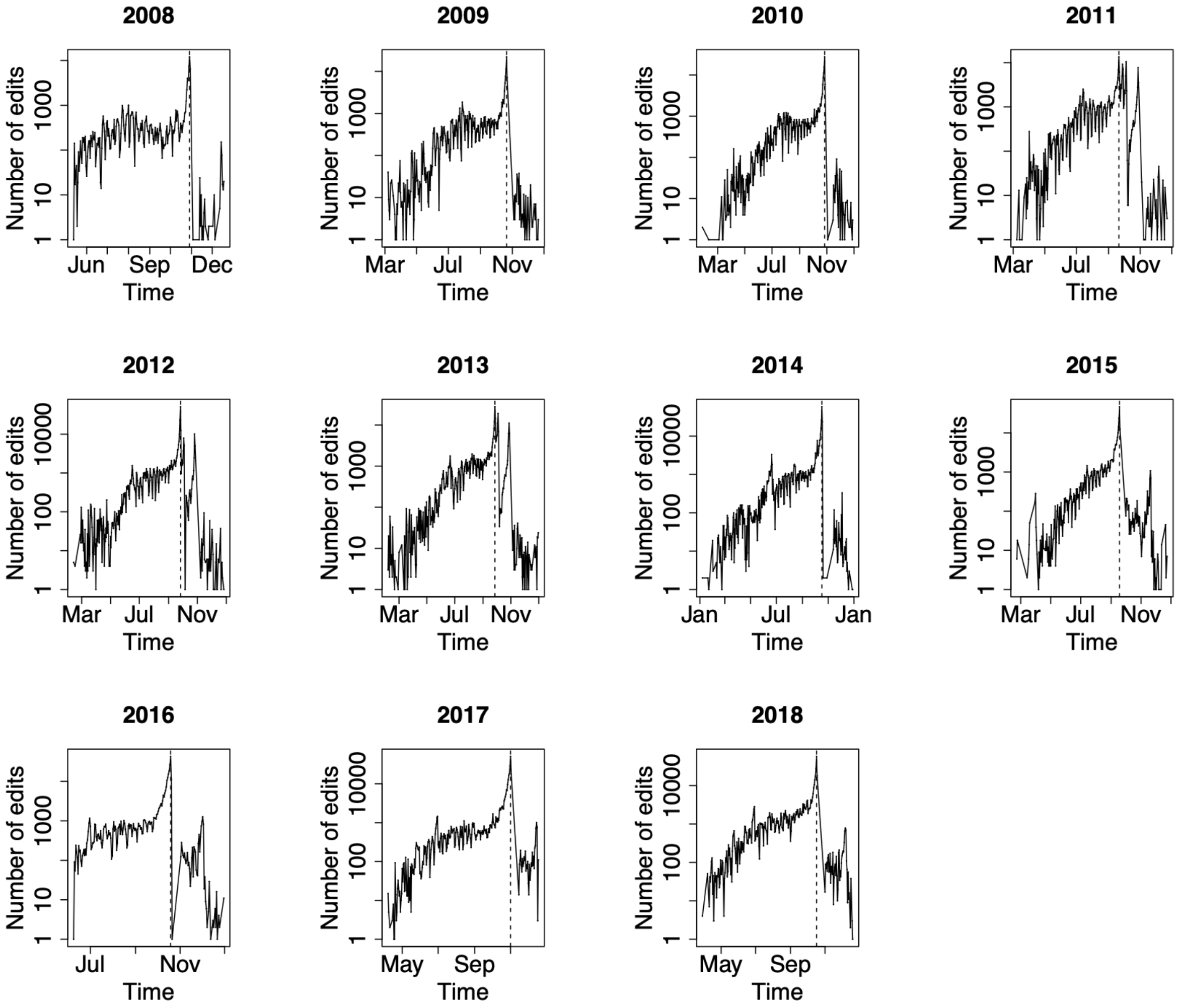

# Figure S11 - Wiki edit activity across years

Total number of wiki edits per day for each competition cycle. Peaks correspond to wiki freeze times. Note the presence of a second peak in 2011, 2012, and 2013, corresponding to a two-phase championship with two wiki freezes. In the main text, we use the peak with maximum edits as wiki freeze (corresponding to the first peak in litigious cases). There are a small number of edits done after wiki freeze time, which correspond to corrections after competition time based on judges feedback.

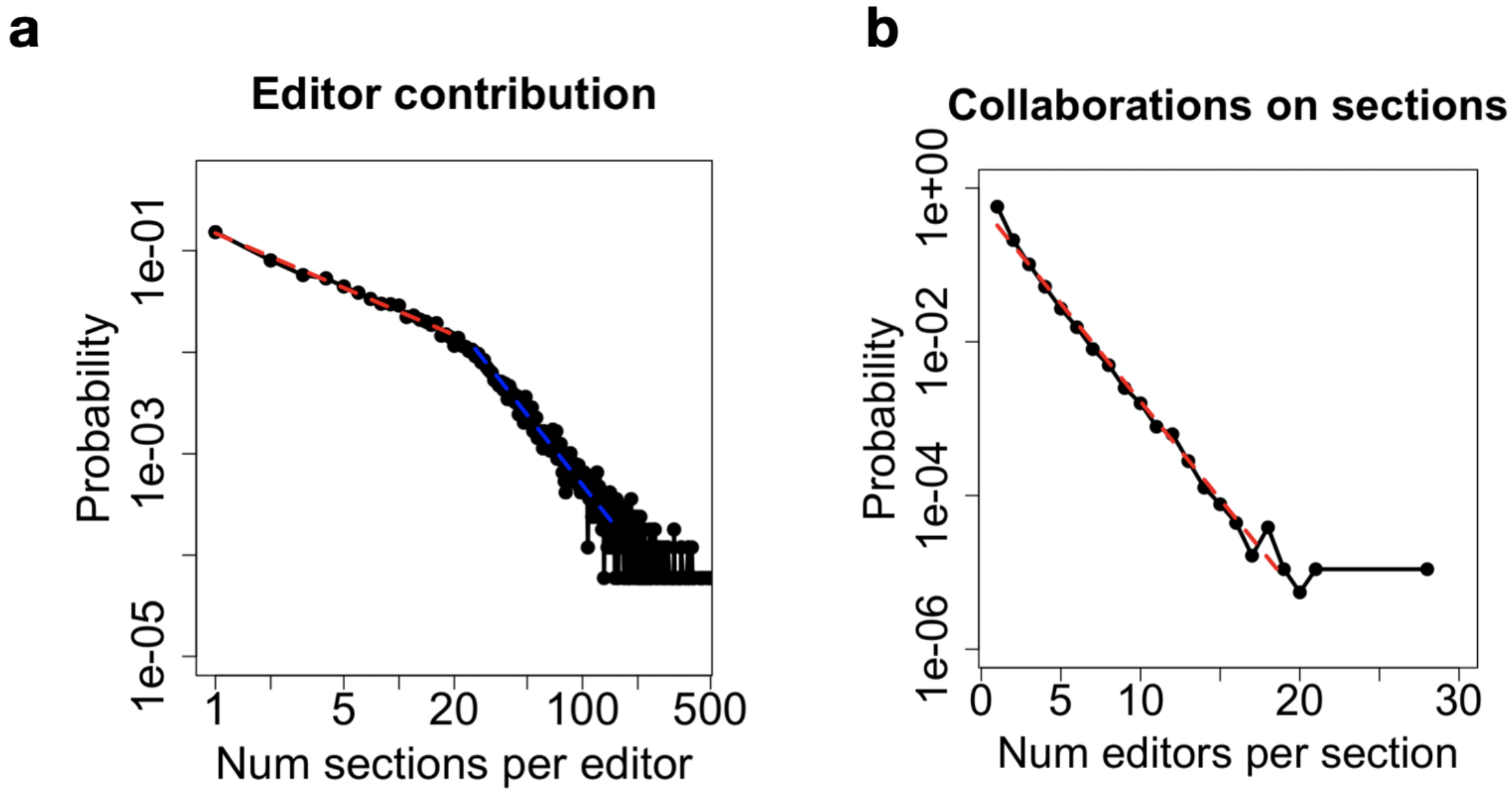

# Figure S12 - Degree distributions of intra-team bipartite network

Same as Fig 3e,f aggregated across all teams.

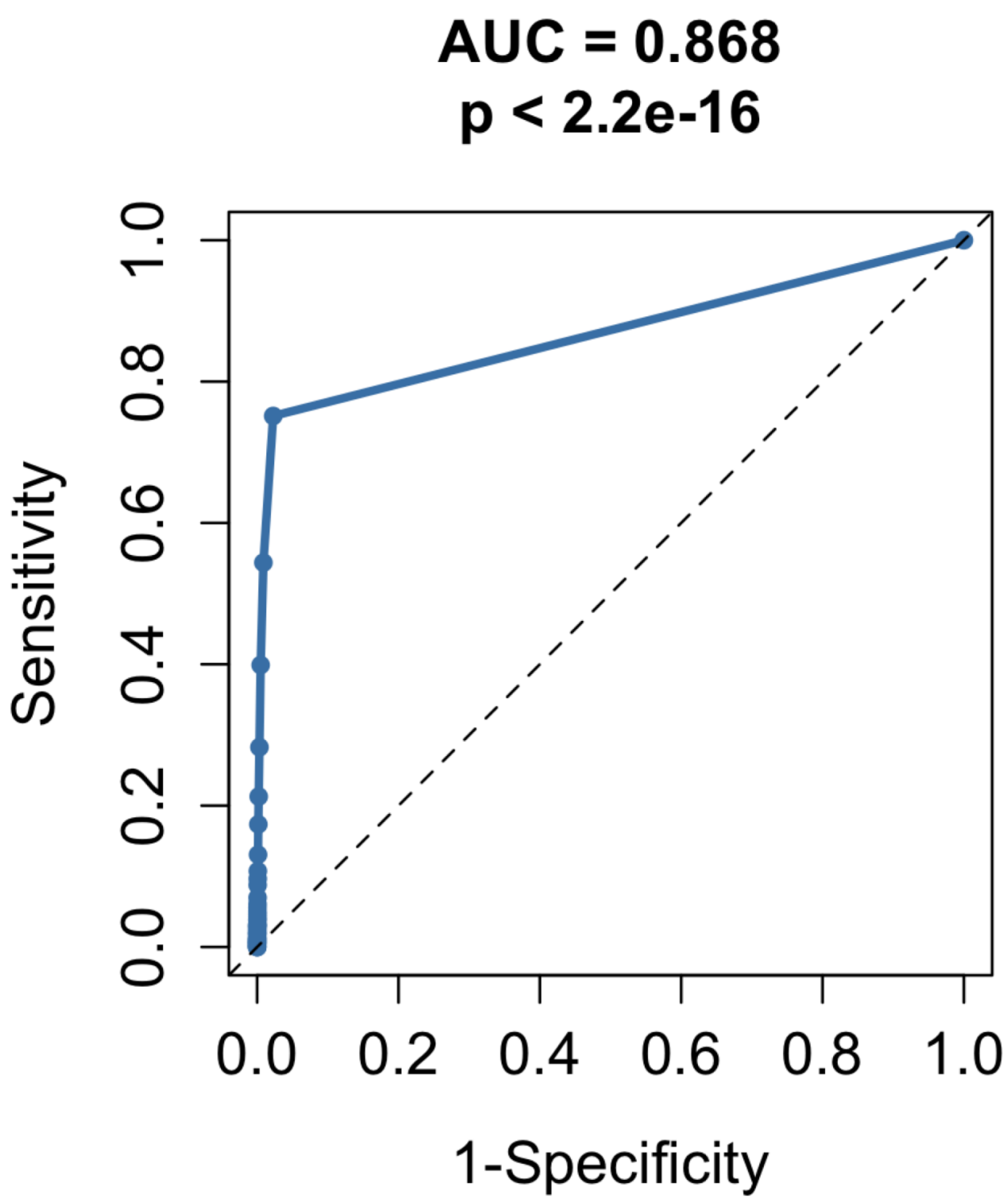

## Figure S13 - Precision recall curves for inter-team mention network

Analysis of the effect of thresholding inter-team edges by the number of mentions on the precision of recovery of manually curated data. ROC curve and corresponding Area Under the ROC curve (AUC) and Wilcoxon p-value for the inter-team network used in main text (Fig 3g).

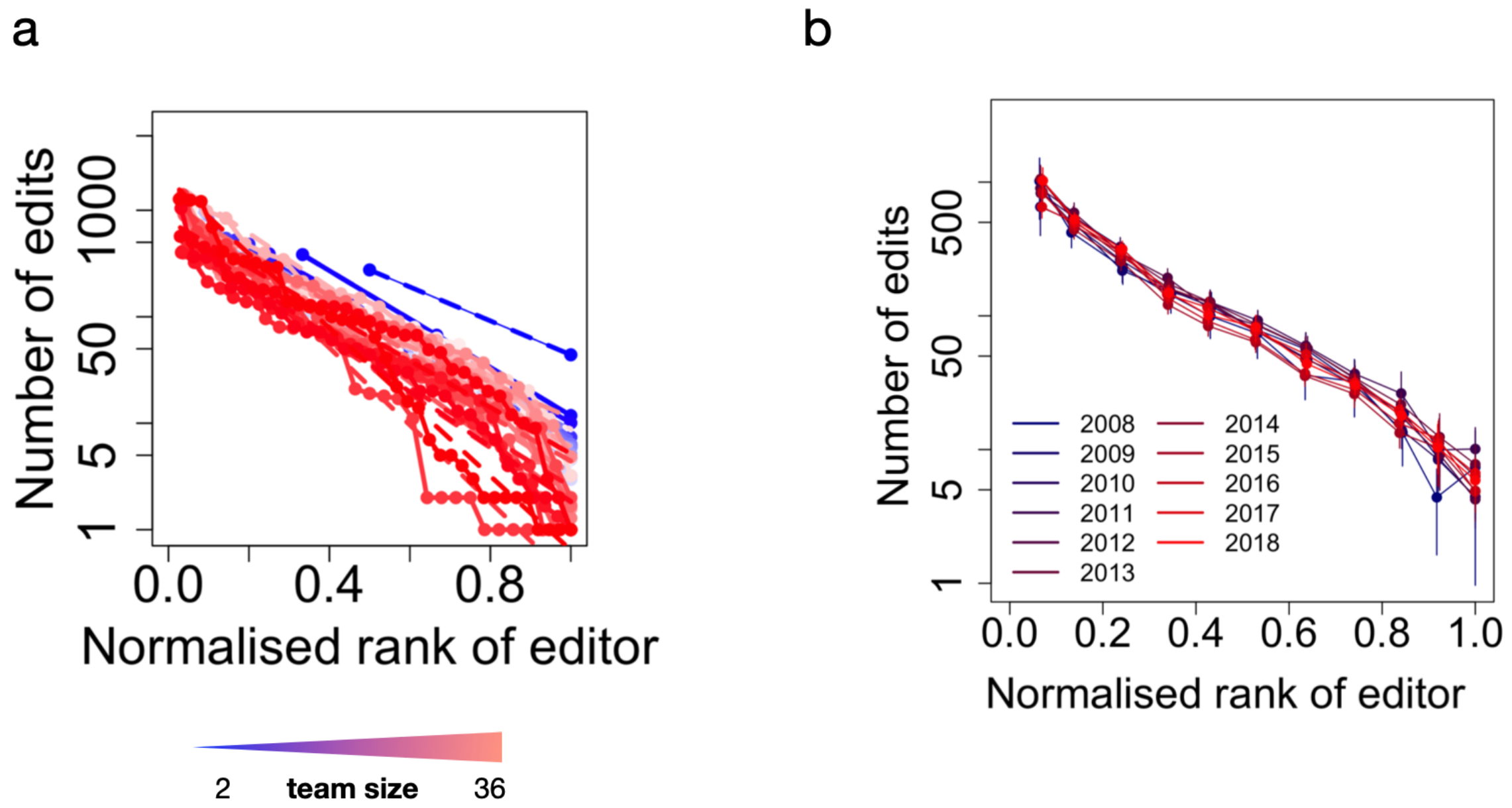

## Figure S14 - Stability of edit inequality across team sizes and across years

Number of edits as a function of normalized rank of editor (same as Fig 4c) across teams of different sizes (a) and across competition cycles (b), consistently following an exponential distribution $N_{edits} \sim e^{-\alpha r}$.

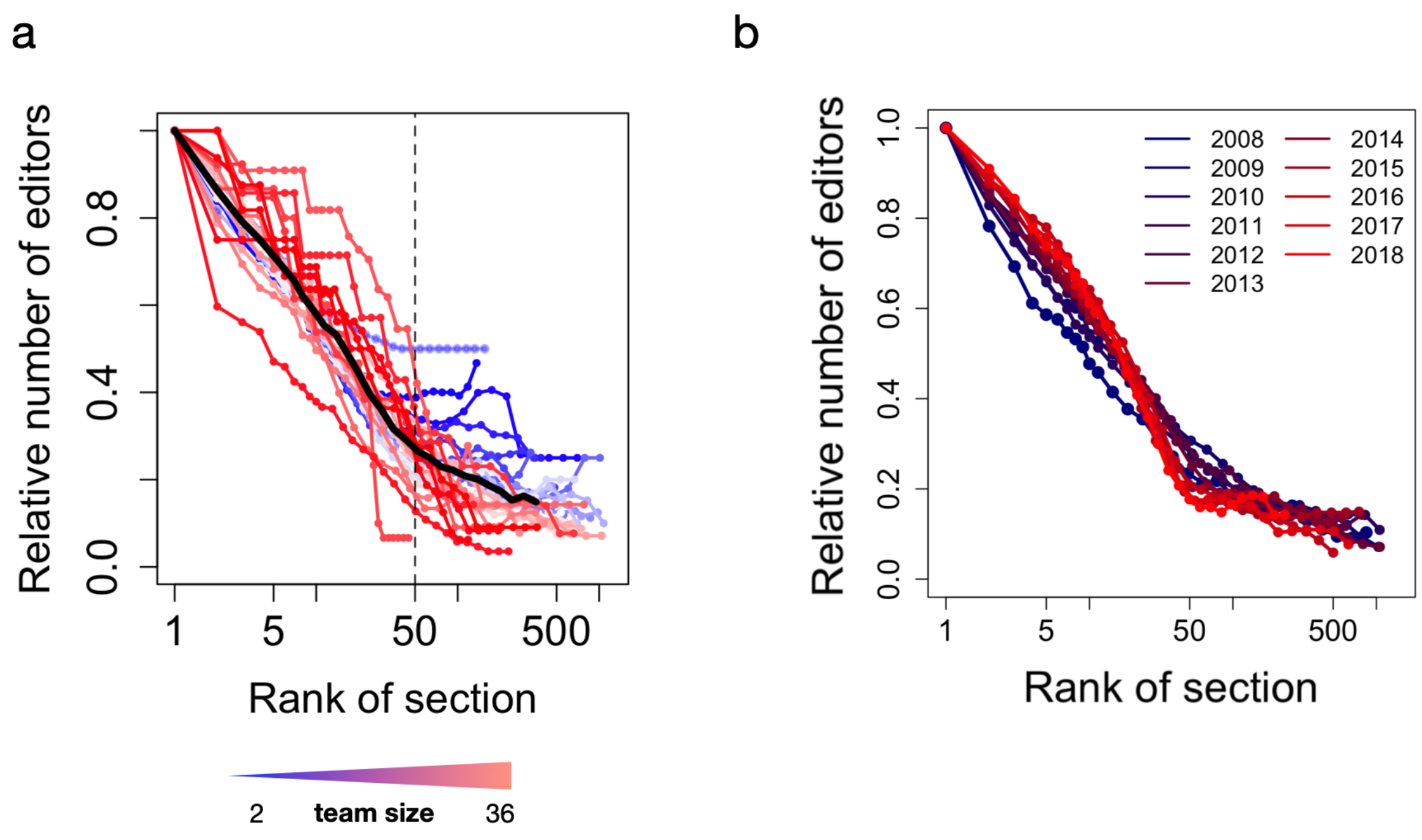

# Figure S15 - Stability of intra-team collaboration across team sizes and across years

Same as figure S13, for the core task overlap.

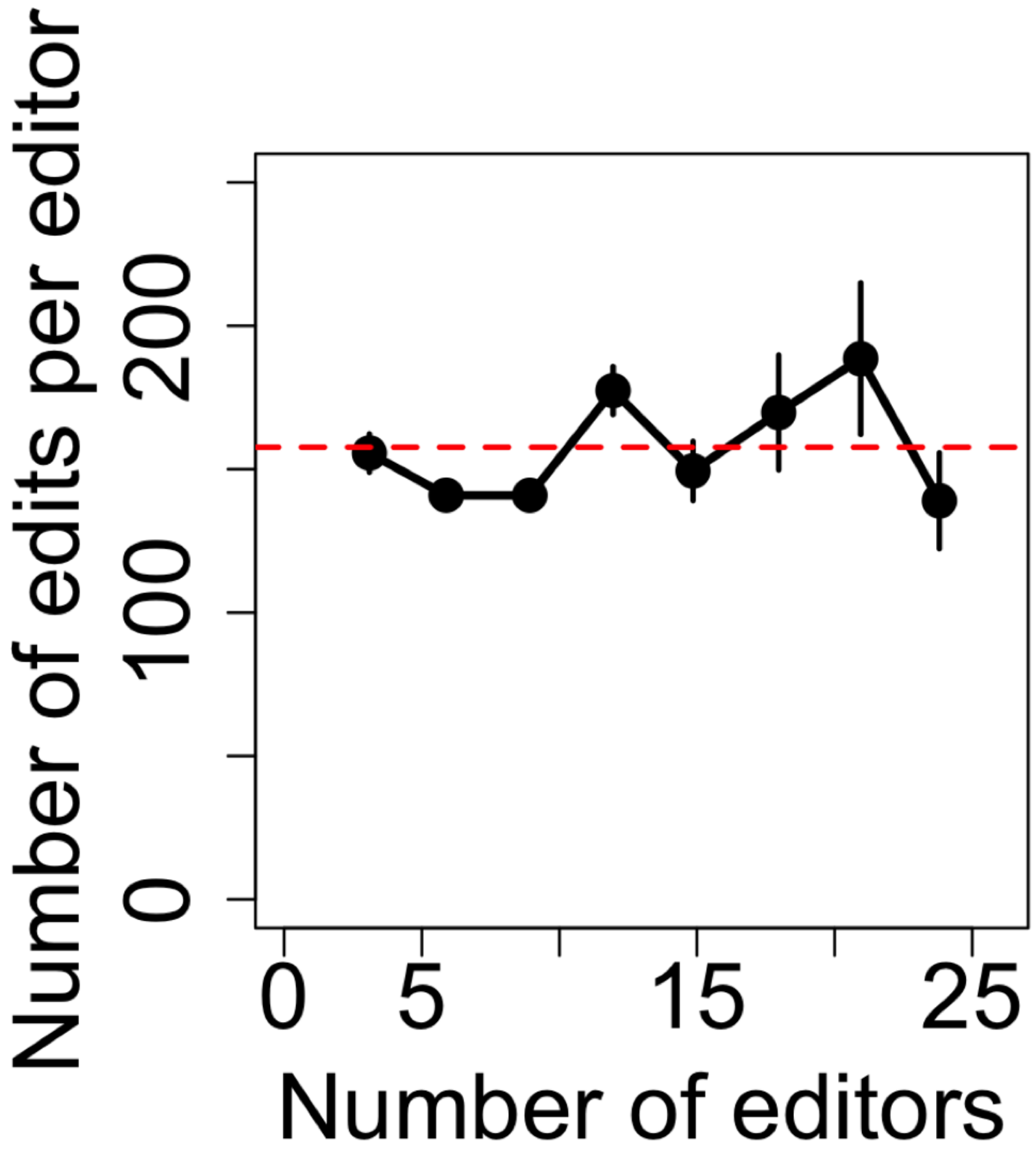

## Figure S16 - Stability of effort per capita across team sizes

Same as Fig 5f, for the number of edits per editor.

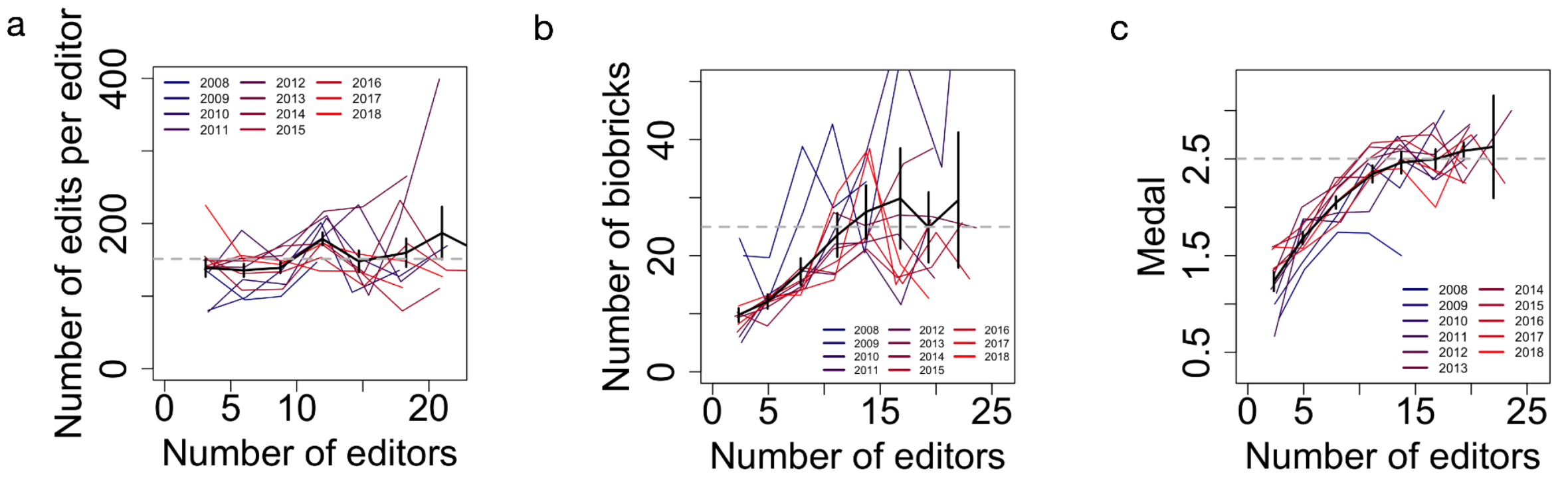

# Figure S17 - Stability of scaling behaviors across years

Same as Fig S16 and Fig 5f,g, per competition cycle.

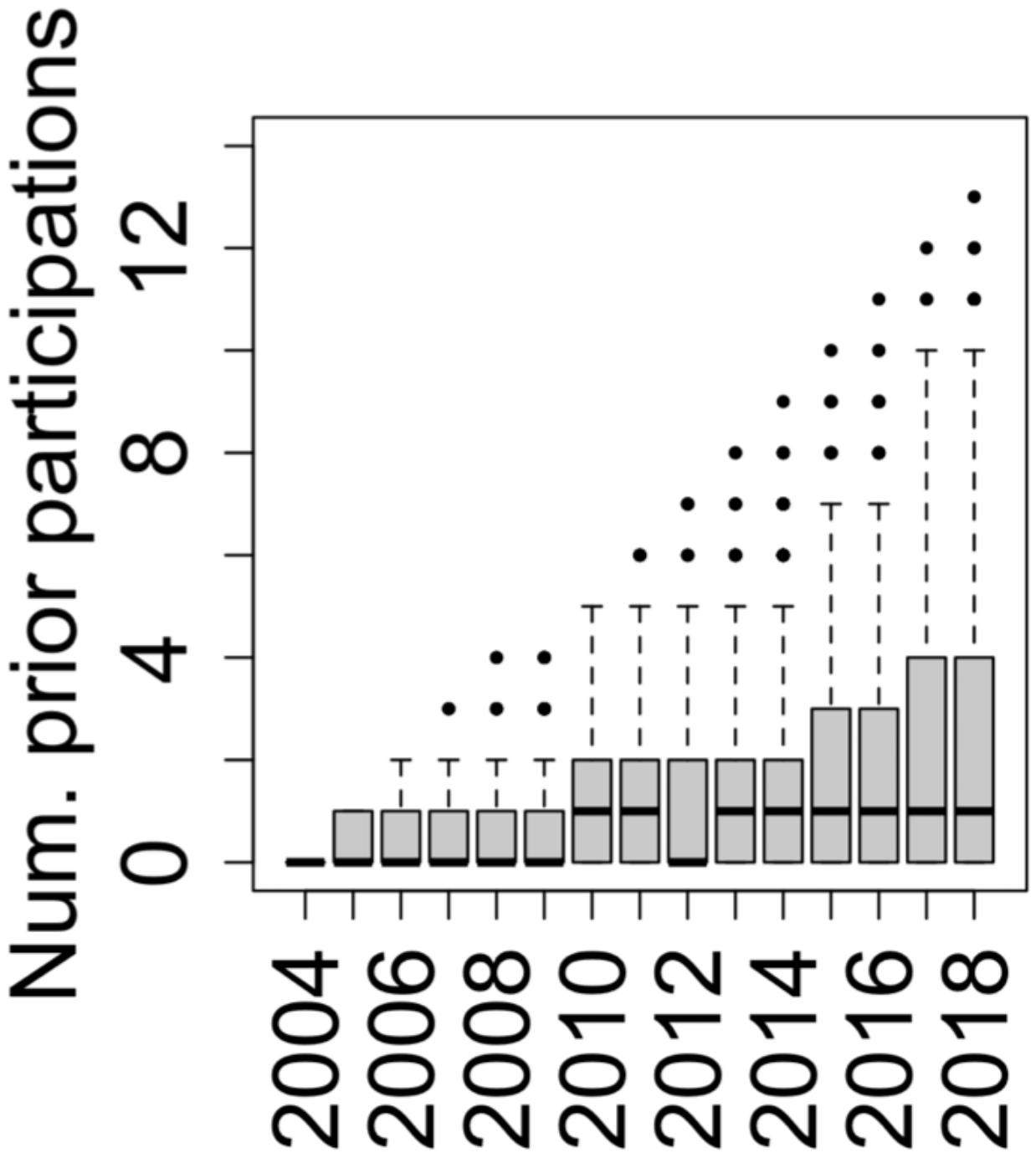

## Figure S18 - Team participations across competitions

Boxplot showing the number of prior participations of teams at each competition cycle.